\documentclass[galaxies,article,accept,moreauthors,pdftex]{Definitions/mdpi} 

\usepackage{pdflscape}	
\usepackage{mathtools}

\def\ltsima{$\; \buildrel < \over \sim \;$}
\def\simlt{\lower.5ex\hbox{\ltsima}}
\def\gtsima{$\; \buildrel > \over \sim \;$}
\def\simgt{\lower.5ex\hbox{\gtsima}}

\newcommand{\Oo}{\displaystyle}

\firstpage{1}
\pubvolume{9}
\issuenum{2}
\articlenumber{29}
\pubyear{2021}
\copyrightyear{2021}
\externaleditor{Academic Editor: {Emilio Elizalde}} 
\datereceived{{02 March 2021}} 
\dateaccepted{{25 April 2021}} 
\datepublished{{date}}
\hreflink{https://doi.org/10.3390/galaxies9020029} 

\usepackage{amssymb}

\Title{Modeling of Spiral Structure in a Multi-Component Milky~Way-Like Galaxy}
\TitleCitation{Modeling of Spiral Structure in a Multi-Component Milky~Way-Like Galaxy}

\Author{{Sergey Khrapov} $^{1,\dagger}$\orcidA{}, Alexander Khoperskov $^{1,\dagger}$\orcidB{}, and Vladimir Korchagin 
$^{2,}$*\orcidC{}}

\AuthorNames{Firstname Lastname, Firstname Lastname and Firstname Lastname}
\AuthorCitation{Khrapov, S.; Khoperskov, A.; Korchagin, V.}
\address{%
$^{1}$ \quad {Volgograd State University}, Universitetsky pr., 100, 400062 Volgograd, Russia; khrapov@volsu.ru (S.K.); khoperskov@volsu.ru (A.K.)\\

$^{2}$ \quad Institute of Physics, Southern Federal University, Stachki Street 124, 344090 Rostov-on-Don, Russia}

\corres{Correspondence: vkorchagin@sfedu.ru}

\firstnote{These authors contributed equally to this work.}

\abstract{Using recent observational data, we construct a set of multi-component equilibrium models of the disk of a Milky Way-like galaxy. The disk dynamics are studied using collisionless-gaseous numerical simulations, 
{{based on the joined integration of the equations of motion for the collision-less particles using direct integration of gravitational interaction and the gaseous SPH-particles.}}
We find that after approximately one Gyr, a prominent central bar is formed having a semi-axis length of about three kpc, together with a multi-armed spiral pattern represented by a superposition of $m=$\,2-, 3-, and 4-armed spirals. The spiral structure and the bar exist for at least 3 Gyr in our simulations. The existence of the Milky Way bar imposes limitations on the density distributions in the subsystems of the Milky Way galaxy. We find that a bar does not form if the radial scale length of the density distribution in the disk exceeds 2.6 kpc. As expected, the bar formation is also suppressed by a compact massive stellar bulge. We also demonstrate that the maximum value in the rotation curve of the disk of the Milky Way galaxy, as found in its central regions, is explained by non-circular motion due to the presence of a bar and its orientation relative to an observer.
 }


\begin{document}


\section{Introduction}

Attempts to understand the phenomenon of spiral structure in galaxies have a long history~\cite{BertinLin1996book,Shu-2016spiral,Bertin+1989ModalApproachI, GerolaSeiden1978Stochastic, Sellwood-etal-2019spiral}. There is a general consensus that the spiral structure is a manifestation of density perturbations propagating in a multi-component stellar-gaseous disk. However, there is not agreement as to the mechanism for spiral generation, one that can successfully explain the rich morphological variety observed in galaxies. To 
describe the observed patterns, a few mechanisms have been suggested. Some researchers treat the spiral structure as the long-lived global modes that last in galactic disks for tens of galactic rotations~\cite{Bertin+1989ModalApproachI,Bertin+1989ModalApproachII,BertinLin1996book}. Others consider the spiral structure to be a transient phenomenon, so that the spiral pattern changes many times during a galaxy's lifetime
\cite{Dobbs-Baba-2014mod, Onghia-2015spiral-halo, Fujii-etal-2018mod, Michikoshi-Kokubo-2016, Sellwood-Carlberg-2019recurrence}.
 The tidal encounter of a small mass companion with a disk galaxy can also be a mechanism for the formation of spiral structure under certain conditions~\citep{PettittEtAll2017StarFormation}. \citet{Khoperskov+2013HaloSpiral} showed that a nonaxisymmetric halo can generate a large-scale spiral density wave in an embedded stellar disk. These authors showed that the pattern is observed 
during many revolutions, even if the disk is gravitationally stable.

Recent observations have provided a significant step in understanding the nature of the spiral structure in galaxies.
Using stellar-cluster catalogs for three nearby spiral galaxies (NGC 1566, M51a, and NGC 628) from the Legacy ExtraGalactic UV Survey, Shabani et al.~\cite{Shabani2018AgeGrad} measured the gradients in stellar ages across the spiral arms within these galaxies. The authors find that the grand design spiral galaxy NGC 1566 shows a significant age gradient across the spiral arms, consistent "with the prediction of a stationary density wave theory". 
The authors attribute the presence of a global mode in this galaxy to a "strong bar''. While an age gradient of the star clusters across the spiral arms in the other two galaxies (M51a and NGC 628) was not confirmed by~\cite{Shabani2018AgeGrad}, the authors do suggest that the spiral structure in M51a is a result of tidal interaction.
Peterken et al. published, almost simultaneously, a paper~\cite{Peterken2019AgeGrad} entitled "A direct test of density wave theory in a grand-design spiral galaxy''. They used the SDSS MaNGA survey to study an age gradient in the galaxy UGC 3825.  They not only determined an age gradient across the arms, but also measured the pattern speed of the spiral arms, finding that it varies little with radius. The authors concluded that their observations are consistent with the presence of a quasi-stationary density wave in UGC 3825. Recently, Bialopetravicius $\&$ Narbutis~\cite{Bialopet2020AgeGrad}
confirmed the existence of an age gradient across the spiral arms in the galaxy M83.
\mbox{Abdeen et al. ~\cite{Abdeen2020AgeGrad}}
looked for an age gradient across the spiral arms in fifteen nearby spiral galaxies and confirmed the existence of a gradient in all of them. Remarkably, Abdeen et al.~\cite{Abdeen2020AgeGrad} found an age gradient in the galaxy NGC 5194 (M51a).
This galaxy is a non-barred galaxy of Sb-type
and, thus, the galaxy shows the presence of a global density 
perturbation not associated with a bar.


Our knowledge of the spiral structure of the Milky Way galaxy is more limited. The morphology of the Milky-Way spiral structure (the number of arms, pitch angle, position(s) of corotation resonance(s)) remains obscured because of our vantage point, i.e., we view the Galactic disk and its spiral structure edge-on 
\cite{Hou-etal-2009spiral, Hou-Han-2014spiral, Xu-Reid-2016spiral, Bland-Hawthorn-Gerhard-2016MW, Shen-Zheng-2020structureMW}.
Difficulties in observing the far side of the Galactic disk, primarily due to extinction, prevent us from obtaining reliable information regarding the morphology and kinematics of nearly half the Milky Way disk. The use of radio interferometry has allowed some initial progress in the determination of the properties of the spiral structure on the opposite side of the Galactic 
disk~\cite{Sanna-etal-2017far-side}.

Masers are the most important source of information for the determination of the morphology of the spiral pattern in the Milky Way galaxy~\cite{Zhang-etal-2019maser-spiral}.
Complementing the maser data, the GAIA astrometric mission~\cite{KatzEtAll2018GaiaDiscKinematics,XuEtAl2018SpiralGaiaMaser,sak-2020,Griv-2020,Gaia-3} has recently provided positions and kinematics for more than a billion stars in the Milky Way. This will enable us to more reliably infer the morphology of the spiral pattern in our own galaxy. It is important to note that the GAIA data on Milky Way spiral structure in the solar neighborhood agrees with the results that are based on VLBI maser observations~\cite{XuEtAl2018SpiralGaiaMaser}.
The third GAIA release~\cite{Gaia-3} will give new impetus in understanding the Milky Way spiral structure.

The structure of the gas flow in the stellar bar area is complex~\cite{Rodriguez-Fernandez-Combes-2008, 
Baba-etal-2010l-v-Features, Sormani-etal-2015bar, Shen-etal-2010bulge, Fujii-etal-2019mod}.  
For this reason, the ability to reproduce the kinematics of 
features in this region is a useful touchstone for verifying theoretical models.
The asymmetry of the central gas layer can be caused either by a lopsided nuclear bar or by infalling gas flow 
\cite{Rodriguez-Fernandez-Combes-2008}.  The observed galactic HI and CO l-v diagrams demonstrate a clumpy morphology, due to the multiphase interstellar medium, gravitational instability at different scales, and star-formation and supernovae 
feedback~\cite{Baba-etal-2010l-v-Features}.
The bulge of our galaxy shows some significant differences with a classical bulge, which formed as a result of major mergers of 
pre-existing disks~\cite{Shen-etal-2010bulge}.
A series of simulations with a huge number (eight billion equal-mass particles) of particles provides the initial conditions for the dynamic models that are required to reproduce the observed characteristics of the Milky Way ~\cite{Fujii-etal-2019mod}.

\citet{Korchagin2016} modeled the dynamics of a two-component stellar-gaseous Milky Way disk using 2D-simulations. These authors found that a three-armed spiral pattern is generated in the Milky Way disk, and its spiral structure is sustained during at least 
3~Gyr.

Here, we present the results of three-dimensional simulations of the dynamics of a Milky Way stellar-gaseous disk. We model the dynamics of three-dimensional multi-component disks using a set of equilibrium models constrained by observational errors and demonstrate that, in most of the models, a multi-armed spiral pattern (a superposition of spirals with different azimuthal wavenumbers) is generated in the disk.

\section{Observational Data}

\subsection{Density Distribution of Stars}

We use the commonly accepted exponential 
density distribution to approximate the surface density distribution of stars in the disk of the Milky Way

\begin{equation}\label{eq:ExpSigmaStar}
\sigma_{*}(r) = \sigma_0\,\exp\{ -r/r_d \} \,,
\end{equation}
\textls[-15]{where $\sigma_0$ is the central surface density and $r_d$ is the radial exponential scale of the stellar disk.}

Using COBE/DIRBE observational data \citet{DrimmelSpergel2001MW} found that the
distribution of stars in the Milky Way disk can be approximated by an exponential function with radial scale length 
$0.28 R_{\odot}$, where $R_{\odot}$ is the distance of the Sun from the center of the Galaxy. Adopting the distance from the Sun to the center of galaxy to be 8\,kpc leads to a galactic disk radial scale length of 2.24\,kpc. Recent estimates of the radial scale length for  Milky Way old stellar populations within $4 < r < 15$\,(kpc) give a value of \mbox{$2.2\pm 0.2$\,kpc~\citep{Bovy+2016GalacticDisk}}. From a study of the density distribution of red clump stars outside the solar circle~\citep{Liu-etal2017OuterDisk}, find that the radial scale length of the old stellar disk has a value of $r_d=2.37\pm 0.02$~kpc. We assume the density distribution of the Milky Way 
old stellar disk varies exponentially with radius, having a scale length of 2.25~kpc.
However, it is necessary to stress that estimates of the radial scale length of the Milky Way disk vary over a rather large range $r_d=$~(2--4)~\mbox{kpc 
\citep{Gerhard2001StructureMW, DrimmelSpergel2001MW, Piffl+2014darkhalo}}, and the references in these papers.
The value of the central surface density of the disk, $\sigma_0$, is very sensitive to uncertainty in the radial scale length $r_d$ (\ref{eq:ExpSigmaStar}). 
This, in turn, imposes limitations on the density distributions of the central bulge and of the halo of the Milky Way disk.

The two key parameters that determine the disk stellar density distribution are the radial scale length of the old stellar disk and the stellar surface density in the solar neighborhood~\cite{Tyurina-2003}.
Estimates that were undertaken by different authors give values varying over a rather wide range. We adopt for the surface density of stars in the solar neighborhood a value of $33.4 \pm 3 \,M_{\odot} / {\rm pc}^2$ from the recent paper by \citet{kee2015}. 
With the these parameters, the total mass of the old stellar disk of the Milky Way is approxiamtely (4--5)$\times  10^{10}\, M_{\odot}$.

 {Observations of edge-on galaxies show that the vertical density distribution of stellar disks can be approximated by functions $\rho_*(z) \propto \exp(-z/h_e)$, $\propto \textrm{sech}(z/h_1)$, $\propto \textrm{sech}^{2}(z/h_*)$ or their combinations~\cite{Mosenkov-etal-2015vert,Comeron-eta-2018vert,Sarkar-Jog-2020vert}, where $h_e$, $h_1$, and $h_*$ are the vertical scale heights.
 The vertical scale height of a stellar disk often depends on radial coordinate, which complicates the situation. In a simplest model of a one-component isothermal disk, the solution is~\cite{Bahcall-1984vert,khop-biz-2010}}
 
\begin{equation}\label{eq:VerticProfileDensity}
\rho_{*}(r,z) = \rho_{*0}(r)\, \textrm{sech}^{2}(z/h_*) \,,
\end{equation}
{where $h_*$ is the vertical scale height of the disk and $\varrho_{*0}$ is the central volume density.}

{Deviations from (\ref{eq:VerticProfileDensity})
are determined by large number of factors, such as an inhomogeneity of disk velocity dispersion in vertical direction, multi-component nature of the disk that reflects its chemical and dynamical evolution, etc. 
An additional factor causing deviation from (\ref{eq:VerticProfileDensity}) is a dark matter halo, as seen from the equilibrium Jeans equation in vertical direction~\cite{Bahcall-1984vert,khop-biz-2010}:}
\begin{equation}\label{eq:vertravnov}
    \frac{d^2\rho}{dz^2} + 2\frac{d\ln c_z}{dz}\frac{d\rho}{dz} - \frac{1}{\rho} \left( \frac{d\rho}{dz} \right)^2 + \frac{4\pi G}{c_z^2} \frac{\sigma}{2z_0}\, \rho \, \left[ \frac{\rho}{\rho_0} + \frac{2z_0}{\sigma} \left\{ \rho_h - \frac{1}{4\pi G r} \frac{\partial V_c^2}{\partial r} \right\} \right] = 0 \,,
\end{equation}
{where $G$ is the gravitational constant, $\rho(z)$ is the volume mass density, $\rho_0=\rho(z=0)$, $\sigma$ is the surface density, $z_0 = \int_0^{\infty} (\rho(z)/\rho_0)\, dz$, $V_c$ is the circular velocity in the plane $z=0$, $\rho_h$ is the local halo density, and $c_z$ is the velocity dispersion of stars in the vertical direction.  
The solution (\ref{eq:VerticProfileDensity}) corresponds to the case $c_z = \textrm{const}$, so that the expression in curly brackets is equal to zero. The initial equilibrium of the disk in our simulations was built using the iteration procedure that is described in 
\cite{khop-biz-2010}, which was modified for presence of a thin gaseous component.}

{\citet{Juric+2008SDSS} find that the three-dimensional distribution of old stellar population of Milky Way can be represented by two subsystems: the thin disk with a radial scale length about 2.5~kpc and a vertical scale height of 300 pc and the thick disk with a scale length and height of 3.6\,kpc and 900~pc, respectively.}

The vertical scale height of the density distribution of the Milky Way disk is another important parameter influencing the gravitational stability of the system and, thus, in determining the morphology of the spiral pattern.
To explore this, in our numerical simulations we vary the vertical scale height of the collisionless disk from 200 to 380~parsec.

The observational data show that the Milky Way scale height increases with radius~\citep{Wan-etal2017LAMOSTIIvertic}.
Nonetheless, we build our equilibrium disk models with constant vertical scale heights.

\subsection{Rotation Curves}

\begin{figure}[H]
\widefigure
	\includegraphics[width=0.95\textwidth]{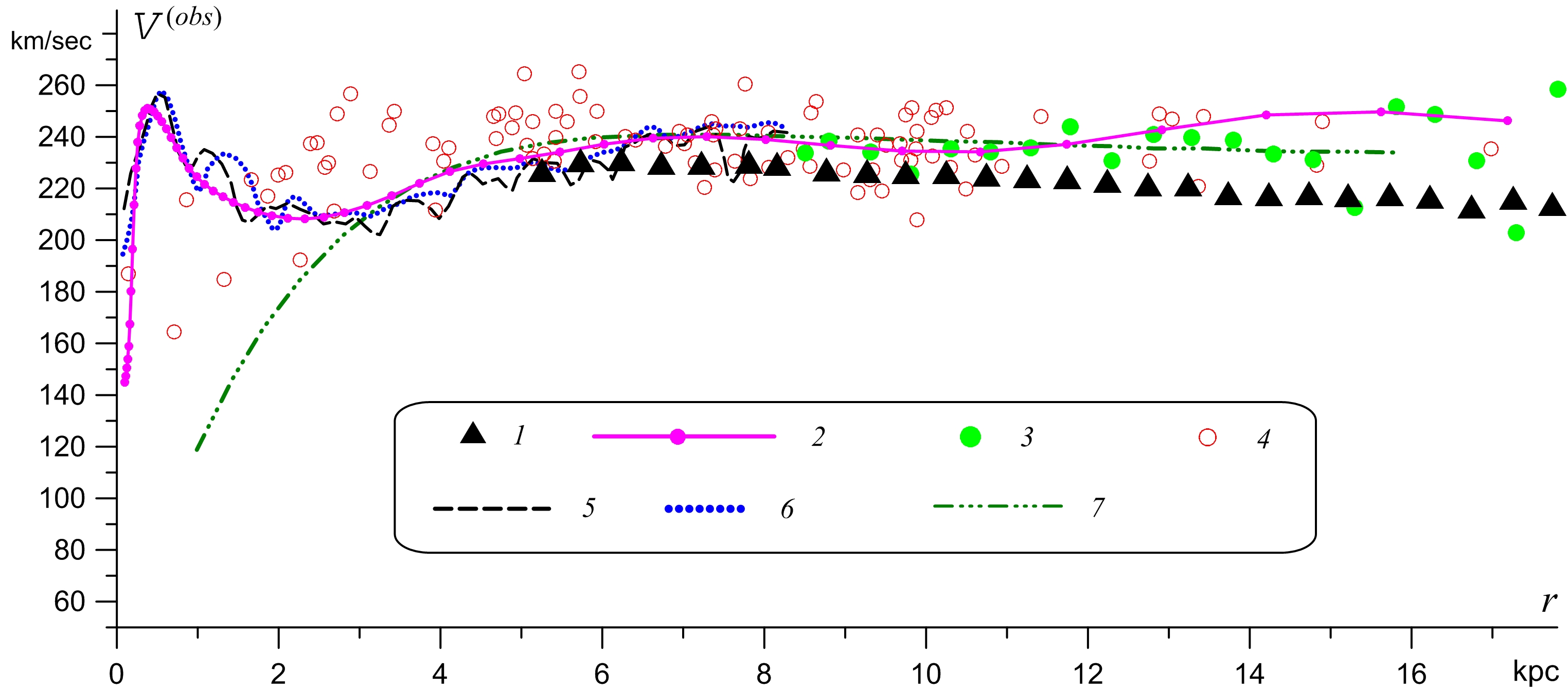}
	\caption{
The rotation curve of the Milky Way galaxy based on different observational data.
The circular velocity $V_c$ and rotation velocity of the gas disk $V_{g}$ are shown: curve ${\textit{1}}$ is the circular velocity from WISE, 2MASS, and Gaia~\cite{Eilers-etal-2019kinem-Gaia}; $2$ is the result of compiling a large amount of data from 
various spectroscopic and trigonometric measurements from radio to optical wavelengths~\cite{Sofue-2020}; $3$ is the reconstructed circular velocity using Gaia Collaboration data and simulations that are based on the Jeans equation
~\cite{Chrobakova-etal-2020Vrot}; $4$---maser observations from~\cite{Rastorguev2017}; gaseous rotation curves inside the solar circle~\citep{Marasco2017}: atomic hydrogen (curve \textit{5}), molecular hydrogen (curve \textit{6}); the maser rotation curve from \citet{Reid2014} is shown by line \textit{7}.
 }
	\label{fig:RotObs}
\end{figure}

\textls[-5]{Figure~\ref{fig:RotObs} shows the observed rotational curve of the Milky Way disk based on various observations.
There is fairly good agreement between the different determinations of the rotation curve in the range for 
$r\simeq$~(6--14)~kpc.
However, the curves differ considerably in the inner regions of the Galaxy.}

\end{paracol}
\nointerlineskip
\begin{paracol}{2}
\switchcolumn

\textls[-5]{New kinematical data that are based on maser observations provide independent and promising data on the kinematical properties of the cold 
component
of the Milky Way \mbox{disk~\citep{Reid2014, Rastorguev2017, Quiroga-Nunez-Reid2017maserSpiralStructr}}. However, despite the fact that 
maser observations directly determine the parallaxes of sources, one has to keep in mind that the reconstruction of the Galactic rotation curve involves a number of assumptions, such as, for example, the model of the Galactic 
potential, which is necessary to reconstruct the rotation curve
\citep{BobylevBajkova2013}.}

The Galactic rotation curve has a local maximum in the inner disk region, as observed in atomic and molecular hydrogen lines 
\citep{Clemens1985VrotMWinner, Sofue2013Vrot, Sofue2017Vrot}. The maximum of the rotational velocity is probably associated with the non-circular motion of gas, as caused by the bar, in the central regions of the Galactic disk.

The angular velocity of the stellar disk is usually lower than that of the gaseous disk. 
Estimates of the rotational velocity of stars in the
solar neighborhood vary from \mbox{200 km\,s$^{-1}$~\citep{Dambis1995Cepheids}} to 237 km\,s$^{-1}$ 
\citep{Lopez-Corredoira2014VrotRedClumpGiants}. A recent estimate by~\citep{Peng-etal2018kinemStellarSun} gives the value of the 
rotational velocity of the old stellar disk (G- and K- dwarfs) as
$V_\odot = 205\pm 20$\,km\,s$^{-1}$.

Outside the stellar disk, the rotation curve of the Milky Way slowly decreases with a radius reaching about 160 km\,s$^{-1}$ at 
a distance of
$\simeq 100$~kpc from the center of the Galaxy~\cite{BajkovaBobylev2016Vrot200kpc}, which apparently points to the presence of the 
extended and massive halo of the Galaxy.

\subsection{Velocity Dispersion of Stars}\label{subDispersionVelocity}

Figure~\ref{fig:DispObs} shows a summary of data on the observational determinations of the radial dependence of the velocity dispersion of stars $c_r$ in the
Milky Way

\begin{figure}[H]
{
	\includegraphics[width=0.7\textwidth]{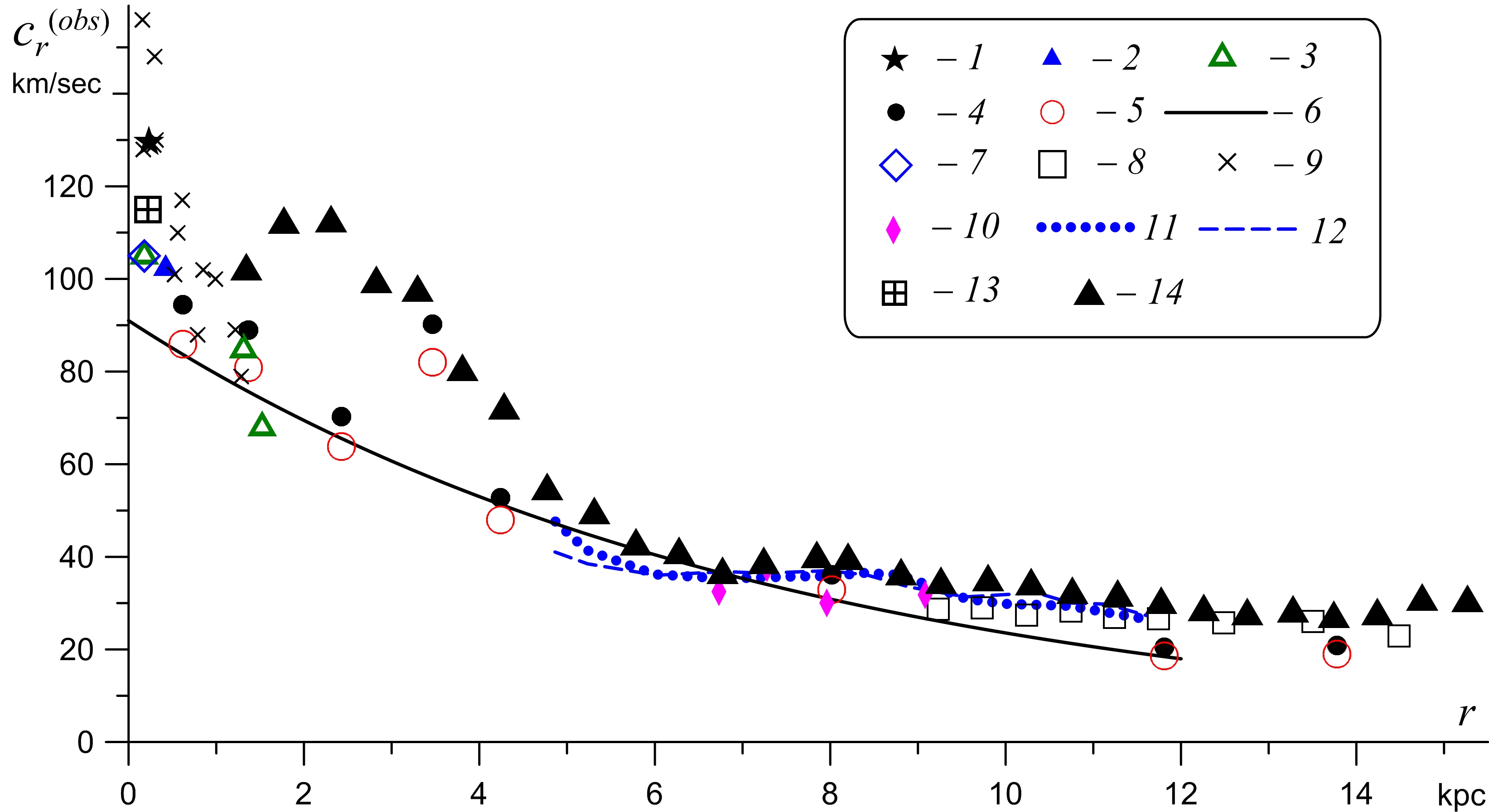} }
	\caption{Radial component of the velocity dispersion of stars as a function of radius. \textit{1}---velocity dispersion of late type giants within 300\,pc radius~\citep{Blum-etal1994}, \textit{2}---\citep{Rich1990DesperCenter}, \textit{3}---
\citep{Minniti-etal1992DispersNearCenter}, \textit{4}---old disk K-giants~\citep{LewisFreeman1989DisperDisc}, \textit{5}---decrease by 10 percents from~\citep{LewisFreeman1989DisperDisc} due to taking into account younger stellar populations, \textit{6}---approximation by $c_r=91$\,[km\,s$^{-1}$]$\exp(r/7.4$\,[kpc]$)$, \textit{7}---\citep{Rich-etal2007DispersCenter}, \textit{8} 
--- outside the solar circle~\citep{Liu-etal2017OuterDisk} taking into account a decrease by 16\% for RGB-stars, $9$---velocity dispersion within bulge region~\citep{Tremaine+2002DispersBulge}, $10$---as based on Gaia-ESO Survey 
\citet{Guiglion-etal2015StellarDispersion}, $11$ and $12$ for giant stars from Gaia in layer $-200$\,pc\,$\le z \le$\,200\,pc for negative and positive azimuths, respectively~\citep{Guiglion-etal2015StellarDispersion}, $13$ is the central velocity dispersion of 
bulge~\citep{Kunder+2012MWbulge}, 14 is the velocity dispersion from Gaia data analysis ~\cite{Eilers-etal-2019kinem-Gaia}.	 }
	\label{fig:DispObs}
\end{figure}

In general, the observational data can be approximated by the function $c_r=91$\,[km\,s$^{-1}$]$/\\ \exp(r/7.4\,[\textrm{kpc}])$. The 
velocity dispersion profile that was obtained by \citet{TiedeTerndrup1999KinemInnerDiskBulge} gives good agreement with the data in the 
disk's central regions of the Galaxy, but in the solar neighborhood with $R_\odot = 8$~kpc the estimate by 
\citet{TiedeTerndrup1999KinemInnerDiskBulge} gives the value
17\,km\,s$^{-1}$, which is too low, and contradicts observations. The velocity dispersion in the central regions of the Milky Way 
disk is about $c_r\simeq$~(110--130)~km\,s$^{-1}$~\citep{Blum-etal1994, Tremaine+2002DispersBulge, Kunder+2012MWbulge} and is likely associated with the bulge stellar populations.

Using the Gaia-ESO spectroscopic survey,
\citep{Guiglion-etal2015StellarDispersion} determined the velocity dispersions $(c_r,c_\varphi, c_z)$ for the thin and thick disk 
separately,
within 6.5\,kpc\,$\le r\le 9$\,kpc. They found, for the thin disk, a velocity dispersion of  $c_r=33$\,km\,s$^{-1}$ (See points 10 in 
Figure~\ref{fig:DispObs}). \mbox{\citet{Piffl+2014darkhalo}} used a sample of 200,000 red giant stars from the Gaia-ESO spectroscopic 
survey and estimated the radial velocity dispersion for the thin disk within 1.5 kpc of the plane to be $c_r=34$\,km\,s$^{-1}$ 
\cite{Piffl+2014darkhalo}.
{It is known that the presence of young stellar populations decreases the effective velocity dispersion in a stellar disk. A detailed study of this effect was conducted  by Rafikov~\cite{Rafikov2001stable},
who discussed the axisymmetric stability criterion of a disk consisting of a few stellar components and a gaseous component.
We take this into account by using a multiplicative factor,
$c_r \rightarrow c_r/1.1$, based on a statistically averaged estimate of the velocity dispersion for the multi-component stellar disk (giants, main sequence stars, white and brown dwarfs):}

\begin{equation}\label{eq:DispersAverage}
    c_r^{(av)} = \left( \sum_l{\sigma_l} c_l^2 \Big/ \sum_l{\sigma_l} \right)^{1/2} \,,
\end{equation}
where $\sigma_l$ and $c_l$ are the surface density and dispersion velocities for the $l$-th component, respectively. Using data of 
\cite{Rafikov2001stable} for nine types of stellar populations, we obtain the above mentioned decrease in the radial velocity dispersion of the stars.

It is known that, for a collisionless disk in equilibrium, the condition

\begin{equation}\label{eq:crcfKappa}
    c_\varphi = c_r\, \frac{\varkappa}{2\Omega} \,
\end{equation}
is satisfied ($\varkappa$ is the epicyclic frequency,  $\Omega$ is the angular speed). We use this condition to determine the initial dependence of the azimuthal 
velocity dispersion along the radial direction.  Our numerical simulations show that this condition is satisfied during the disk~evolution.

Different authors provide values of $0.45- 0.6$ for the ratio $c_z/c_r$ in the solar neighborhood. A recent estimate by
\citep{Peng-etal2018kinemStellarSun} of the velocity dispersions $c_z = 16$\,km\,s$^{-1}$, 
$c_r = 35$\,km\,s$^{-1}$, gives a value for
$c_z/c_r$ of 0.46, which is what we used to build our equilibrium models. The ratio
$c_z/c_r=16$\,km\,s$^{-1}$/35\,km\,s$^{-1}$ is equal to 0.46, based on SCUSS and SDSS observational data for  G and K dwarfs of the thin disk~\citep{Peng-etal2018kinemStellarSun}. 
Estimates by\mbox{ \citet{KatzEtAll2018GaiaDiscKinematics}} and 
\mbox{\citet{AdibekyanEtAll2013Kinemat}} give $c_z/c_r=0.5$ for this ratio.
We choose a value of 0.5 in our simulations for the ratio $c_z/c_r$. 
We note that the kinematical properties of the 
stellar disk change with time during disk evolution, due to disk heating \citet{HaydenEtAll2017GrowthVertDispersionTime}.

\subsection{Stellar Bulge and Bar}

A number of observational studies demonstrate that the Milky Way has a boxy/peanut-shaped bulge. Recent evidence points to the presence of an additional spheroidal
component~\citep{GonzalezGadotti2016MWreview}. The structural properties and origin of the bulge are the subject of intense study. 
In this paper, we neglect the possible evolution of the bulge and consider it to be a fixed structure of the Galaxy contributing to the total gravitational potential. Estimates
of the mass of the Milky Way bulge vary over a rather wide range, (0.9--2.0) $\times \,10^{10}\, M_{\odot}$~\citep{Valenti2016, Portail+2017bulge}, which, in part, is a result of different assumptions regarding the stellar density distribution of the bulge. Estimates of the Galactic bulge's mass are based on photometric and kinematical data. A recent estimate for the bulge, based on photometric data~\citep{Valenti2016}, gives the total mass of stars and 
remnants in the bulge as 
$2.0 \pm 0.3 \times 10^{10}\, M_{\odot}$, which is consistent with the estimate by \citet{Portail+2017bulge}, who derive a total dynamical mass for the bulge equal of $1.85 \pm 0.05 \times 10^{10}\,M_{\odot}$.
The kinematical estimates of the bulge's mass are based on the assumption that the maximum of the rotational velocity of gas close to the galactic center at 0.5\,kpc is caused only by the galactic bulge.  This gives a total bulge mass of approximately
$M_b=0.99\times 10^{10}$~M$_\odot$~\citep{BobylevBajkova2017Bulge}. We vary the total mass of the bulge in our simulations within
(0.7--1.6)$\times 10^{10}\, M_{\odot}$.

The presence of a prominent stellar bar with a large semi-axis of about 
(3.1--3.5)\,kpc and with an axial ratio of about 1:0.4 is another peculiarity of the Milky Way galaxy~\citep{GonzalezGadotti2016MWreview}. The major semi-axis of the bar deviates by approximately $23^\circ$ from the direction to the Galactic center~\citep{Michtchenko2018MWfullModelbar}. 
A successful dynamical model of the Milky Way must agree with the available observational data and reproduce the observed parameters of the bar.

\subsection{Gas Distribution}

Gas plays a significant, possibly crucial, role in forming and sustaining the spiral structure in galaxies.  
\citet{GoldreichLynden-Bell1965II}
pointed to the importance of gas in the formation of spiral structure: S0 galaxies are topographically similar to normal spirals, but they have no gas, no dust, and no spiral arms.
This suggests that stellar dynamics is not solely responsible for arm formation.

The total amount of gas within 30~kpc of the Milky Way disk is approoximately $8 \times 10^{9} M_{\odot}$~\citep{NakanishiSofue2016}. In our simulations, we vary the total amount of the disk's gaseous component
from $3.5 \times 10^{9} M_{\odot}$ to $6.5 \times 10^{9} M_{\odot}$ within the
optical radius of the disk,  $R_{opt}=4r_d=9$\,kpc, s listed in Table \ref{tab:ParametersModels}. As a result, the observationally limited ratio of masses of gas and stars in the Milky Way disk varies within $\mu_g=M_g/M_d =\,$(0.13--0.17). In our simulations, the gas surface density in the solar neighborhood was kept within the observed estimates of ($H_2$, $HI$ and $HII$) $\sigma_{g\odot}=13.7\pm 1.6\,M_\odot$\,pc$^{-2}$~\citep{kee2015}.
 The observational data for distant galaxies indicate a significant role of gas accretion, minor and major mergers, which makes the stellar-gas disk a non-conservative system, exerting a strong influence on the kinematics and morphology of galaxies 
\citep{KhoperskovS-etal-2021}. 

{
The current computational possibilities allow following the dynamics of Milky Way-like galaxy for more than 10 Gyr.
Recent cosmological simulations undertaken by a few groups for following the formation of galaxies in isolated Milky Way-mass are impressive \mbox{haloes 
\citep{Graziani-etal-2015init-state, Graziani-etal-2017init-state, Grand-etal-2017, Hopkins-etal-2018}}.
These simulations follow many aspects of galaxy formation, including black hole accretion and its feedback, feedback from massive stars, stellar and chemical evolution, metallicity-dependent cooling, star formation, and an influence of magnetic fields, e.g., mock images of the Milky Way-like galaxy in FIRE-2 simulations demonstrate, at z =0, a developed multi-armed spiral pattern. Our simulations use a more simplified model when gas is treated as a quasi-isothermal one, but we base the simulations of disk dynamics on  the current observational knowledge of Milky Way stellar and gaseous rotation curves, stellar, and gaseous velocity dispersions, the density distributions of gaseous and stellar components in the Milky Way disk, as well as  on the knowledge of size and density distribution in the stellar bulge of the Milky Way galaxy.
 }

\subsection{Dark Matter Halo}

The equilibrium of the galactic disks depends on the dark mass distribution in \mbox{halos 
\cite{zasov-etal2017, Lucia-2019dark}}.
The rotation of the Galactic disk is determined by the common gravity of the disk itself, the bulge, and of the massive dark matter halo. Once the rotation curve is specified, as well as the mass and density distributions of the disk and bulge, one can determine the mass of the dark-matter spherical halo. Within the optical radius of the Milky Way disk, the halo mass exceeds that of the disk (see Table \ref{tab:ParametersModels}). 
The Galactic halo reaches a mass of approximately (5.5--10.3)$\times 10^{10}\, M_{\odot}$ within $R_{opt}$. In numerical 
models, with the value of $R_{opt}$ differing depending on the adopted radial scale length of the disk $r_d$ (See Table \ref{tab:ParametersModels}). Based on 
kinematical data for ~200,000 giant stars,
\citep{Piffl+2014darkhalo} estimated the mass of the dark-matter halo within the solar radius to be $M_h(r\le R_{\odot})=(6\pm 0.9)\times 10^{10}\,M_\odot$. {However, the galactic halo extends to much larger distances than the optical radius of the stellar disk (for a review see~\citep{zasov-etal2017, Deason-etal-2021-100kpc} and the references there).}

\section{Numerical Code and Stability Criteria}

\subsection{Equations and Numerical Algorithm}

We treat the dynamics of the 3D stellar-gaseous galactic disk self-consistently. The gaseous subsystem is modeled by $N_g$ Smoothed Particle Hydrodynamics (SPH) particles. 
The dynamics of the collisionless (stellar) disk is modeled using $N_*$ particles and a direct particle-particle integration scheme. We set the numbers of stellar and gaseous particles equal, $N_g = N_* = N/2$, where $N=N_g+N_*$ is the total number of particles used in the numerical simulations.

{ Because the star formation is not taken into account in our model, the relative number of the gaseous and stellar particles does not influence the results of simulations. Therefore, based on the convenience of modeling, we chose $N_* = N_g$.
}

The equations of motion of gravitationally interacting SPH and collisionless disk particles in the external gravitational fields of an axisymmetric isothermal halo ${\bf f}^{h}$ and bulge ${\bf f}^b$ are as follows:

\begin{equation}\label{Eq-dv_dt}
\frac{d^2 {\bf r}_i}{d t^2} = \begin{cases}
\displaystyle - \frac{\nabla p_i }{\varrho_i }  + {\bf f}^{h}_i + {\bf f}^b_i + \sum_{j=1,j\neq i}^N{{\bf f}_{ij}}   ,&0 \le    i   
< N_g; \\
\displaystyle  {\bf f}^{h}_i + {\bf f}^b_i + \sum_{j=1,j\neq i}^N{{\bf f}_{ij}} , &N_g \le    i   < N;
\end{cases}\,,
\end{equation}
where $t$ is the time, $\nabla$ is the Hamilton operator,
the radius-vector ${\bf r}_i(t)$ specifies the position of the $i$-th particle in space, while $\varrho_i$ and  $p_i$ are the mass density and gas pressure of the $i$-th particle, respectively.
{We use the analytical halo model described in the works~\cite{{k-z-t2003num, Khoperskov+2012ngc5247}}.
}
The gravitational interaction between the $i$-th and $j$-th particles is given by the equation:

\begin{equation}\label{Eq-self-gravity}
{\bf f}_{ij} = -G  {\frac{ {m_j} \,({\bf r}_i - {\bf r}_j)}{|({\bf r}_i - {\bf r}_j)^2 + \delta^2|^{3/2}}}\,,
\end{equation}
where $m_j$ and ${\bf r}_j$ are the mass and radius-vector of the $j$-th particle, respectively, and 
$\delta$ is the gravitational softening length, preventing unrealistic accelerations during close encounters of the particles~\cite{Smirnov-Sotnikova2017,  Maureira-Fredes-Amaro-Seoane2018soft}. 
In our simulations, the number of particles is
$N = (2^{19}$--$2^{21})$, and the softening length is taken to be $\delta \simeq 40$~pc.

{With the maximum resolution $N_*=N_g=2^{20}$ in simulations listed in Table A1, masses of the stellar and the gaseous particles are within $m_* \sim 2$--$3\times 10^4$\,M$_\odot$, $m_g \sim 2.4$--$6.5\times 10^3$\,M$_\odot$.}

In order to describe the dynamics of the gaseous disk, one must supplement the equations of motion with the equation of specific internal energy conservation, $e_i$, and with the equation of state of the gas $e = e(p, \varrho)$:

\begin{equation}\label{Eq-de_dt}
\frac{d e_i}{d t} = - \frac{p_i}{\varrho_i} \nabla \cdot {\bf v}_i\,,
\end{equation}
\begin{equation}\label{Eq-e_p_rho}
e_i =  \frac{p_i}{(\gamma-1)\varrho_i}\,,
\end{equation}
where ${\bf v}_i$ is the velocity vector of the $i$-th particle and $\gamma$ is the adiabatic index. 
We choose an equation of state for the gas close to an isothermal one with 
$\gamma = 1.05$ to decrease the effect of gas heating by shock fronts during the evolution of the gaseous disk.
Such an approach is employed as the simplest cooling model in various studies of astrophysical gas \mbox{dynamics
\cite{Sawada1986,Bisikalo,SmallGamma1,SmallGamma2}}.
A more realistic consideration of thermal processes requires a cooling function that depends on temperature, density, {and gas metallicity}
\cite{Vasiliev2012,sak2013,Vorobyov,vasiliev2017}.
A$\,$correct description of gas cooling is possible for multicomponent gas models that take chemical reactions {and gas metallicity} into account, 
as is done in the modeling of the molecular clouds in the disks of spiral galaxies
\cite{dobbs2013,Khoperskov-Vasiliev2016,dobbs2016}.
An accurate accounting of gas cooling should also include radiation transfer (see, e.g.,~\cite{sak2013}), and we leave this for further consideration.

In accordance with the SPH-approach~\citep{Monaghan1992}, the density of gas that is associated with the i-th gas particle, the equation of motion (\ref{Eq-dv_dt}), and the energy conservation Equation (\ref{Eq-de_dt}) can be written in the following form 
(details can be found in 
\citep{KhrapovKhoperskov2017SPH}):
\begin{equation}\label{Eq-SPH-mass-density}
\varrho_i = \varrho ({\bf r}_i) = \sum_{j=1}^{N_g} m_j \, W(|{\bf r}_i-{\bf r}_j|,h_{ij}) \,.
\end{equation}
\begin{equation}\label{Eq-dv_dt-SPH}
\frac{d {\bf v}_i}{d t} = - \sum_{j=1,j\neq i}^{N_g} {m_j\,\Pi_{ij}\, \nabla W_p\left(|{\bf r}_i-{\bf r}_j|,h_{ij}\right)}
+ {\bf f}^{h}_i + {\bf f}^b_i + \sum_{j=1,j\neq i}^N{{\bf f}_{ij}}\,,
\end{equation}
\begin{equation}\label{Eq-de_dt-SPH}
\frac{d e_i}{d t} = \frac{1}{2}\,\sum_{j=1,j\neq i}^{N_g} {m_j\,\Pi_{ij}\, ({\bf v}_i-{\bf v}_j) \cdot \nabla 
W_p\left(|{\bf r}_i-{\bf r}_j|,h_{ij}\right)}
\,.
\end{equation}

Here, $W$ is the Monaghan smoothing kernel~\citep{Monaghan1992} and  $W_p$ is the smoothing kernel that is used for the approximation of pressure forces~\citep{Muller+2003SPH, KhrapovKhoperskov2017SPH}, and $h_{ij}=0.5\,(h_i+h_j)$ is the effective smoothing length, where the smoothing length for each particle depends on its mass and density as $h_i = 1.3 \left( m_i / \varrho_i\right)^{1/3}$ 
\citep{Monaghan2005SPH, KhrapovKhoperskov2017SPH}.  $\displaystyle \Pi_{ij} = \frac{p_i}{\varrho_i^2} + \frac{p_j}{\varrho_j^2} + \nu_{ij}^a$ is the symmetric approximation of the pressure forces, and  $\nu_{ij}^a$ is the artificial viscosity 
\citep{KhrapovKhoperskov2017SPH}.
{We have, for the artificial viscosity, $\nu_{ij}^a$:}

$$
\nu_{ij}^a = \frac{\mu_{ij}\,(\beta \, \mu_{ij} - \alpha \, c_{ij})}{\varrho_{ij}}\,, \quad \textrm{where} \quad
\mu_{ij} =
\begin{dcases*}
\frac{h_{ij}\,\Delta\mathbf{r}_{ij} \cdot \Delta\mathbf{v}_{ij}}{|\Delta\mathbf{r}_{ij}|^2 + \eta \, h^2_{ij}}     
,&$\Delta\mathbf{r}_{ij} \cdot \Delta\mathbf{v}_{ij} <0$ \\
0                                   ,&$\Delta\mathbf{r}_{ij} \cdot \Delta\mathbf{v}_{ij} \ge 0$
\end{dcases*}
\,,
$$
{$\Delta\mathbf{r}_{ij} = \mathbf{r}_i - \mathbf{r}_j$, $\Delta\mathbf{v}_{ij} = \mathbf{v}_i - \mathbf{v}_j$, $\varrho_{ij} = 0.5 
\left(\varrho_i + \varrho_j\right)$, $c_{ij} = 0.5 \left(\sqrt{\gamma p_i/\varrho_i} + \sqrt{\gamma p_j/\varrho_j}\right)$.
The empirical constants $\alpha$, $\beta$, and $\eta$ determine the value of the artificial viscosity. In our simulations ee used the values $\alpha=0.5$, $\beta=1$, and $\eta=0.1$.}

{If the smoothing kernel W (Monaghan cubic spline ~\citep{Monaghan1992})
is used to calculate the pressure gradient, a nonphysical numerical clustering of particles will occur in high-pressure regions~\citep{Desbrun1996}. This is caused by the weakening of the interaction between the particles in the vicinity of
$\Oo 0 < \xi < \frac 2 3$ and $\Oo \lim_{\xi \rightarrow 0} \frac{\partial W}{\partial \xi}=0$, where $\xi=|\mathbf{r}_i - 
\mathbf{r}_j|\,/\,h_{ij}$ is the distance between the i-th and j-th particles. To eliminate this and improve the stability 
of the numerical algorithm, we use the smoothing kernel $W_p$ taken from~\citep{Muller+2003SPH}:}

\begin{equation}\label{Eq-SPH-Kernel-p}
W_p(\xi,h)   =   \frac{15}{64 \pi h^3}
\begin{dcases*}
(2 - \xi)^3                 ,&$0 \le      \xi   \le 2$; \\
0                                   ,&$ \xi \ge 2$.
\end{dcases*}
\end{equation}

{From Equation (\ref{Eq-SPH-Kernel-p}), it follows that $\Oo \lim_{\xi \rightarrow 0} \frac{\partial W_p}{\partial 
\xi}=-\frac{45}{64 \pi h^4} \neq 0$.}

For the numerical integration of the differential Equations
(\ref{Eq-dv_dt-SPH}) and (\ref{Eq-de_dt-SPH}), we use the predictor--corrector scheme of second-order accuracy 
(the so-called leapfrog method). We use the direct particle-particle algorithm to calculate the gravitational forces.
The leapfrog method allows for us to simulate the dynamics of the disk systems, even in the cartesian coordinate system. 
\citet{Khrapov+2018Chel} have shown that use of the leapfrog method in double-precision simulations conserves the total energy, momentum, and angular momentum of the equilibrium system with $N = 2^{20}$ particles with an accuracy of $10^{-5}$, $10^{-15}$ and $10^{-13}$, respectively. For single-precision simulations, the accuracy of conservation of the above-mentioned quantities is equal 
to $10^{-3}$, $10^{-2}$ and $10^{-3}$, respectively \citep[see][]{Khrapov+2018Chel}.

Details of the realization of the predictor-corrector method are described by~\citep{KhrapovKhoperskov2017SPH,Khrapov+2018Chel}. Here, we outline the coordination procedure for simulations of gaseous and stellar disks. For the gaseous disk, the integration time step $\Delta t_g(t)$ is limited by the stability condition of the SPH-algorithm 
\citep[see e.g.][]{KhrapovKhoperskov2017SPH}, while, for the stellar disk, the time step is limited by the condition of applicability of Newtonian gravity, namely the time step in the collisionless simulations should be greater than the time of light propagation in the region of the simulations $\Delta t_* \ge \Delta t_{crit}$. Here, $\Delta t_{crit}$ is the propagation time of light within the region. If we choose $\Delta t_* = \Delta t_g$, the angular momentum conservation is satisfied with an accuracy of $10^{-13}$. However, in this 
case, the condition of applicability of Newtonian gravity fails, and the total integration time of the problem increases by factors of tens to hundreds.  Therefore, we choose the value $\Delta t_* = \Delta t_{crit} \simeq 2 \times 10^5$ years in our simulations. The calculation of the gravitational interaction between all of the particles was carried out once for a time interval $(t,\,t+\Delta t_*)$ using the expression (\ref{Eq-self-gravity}). For the gaseous disk, the number of time steps is large ($n_g \gg 1$) for each time interval $(t,\,t+\Delta t_*)$, in order that the gravitational force vector is constant during each time interval, which leads to an accuracy of conservation of angular momentum of $10^{-2}$. The following correction procedure for the velocities of gaseous particles at the last time step $\Delta t_g(t_{n_g})$ allows us to increase the accuracy of angular momentum conservation to $10^{-8}$:

\begin{equation}\label{Eq-Vx_g_corr}
v_{x}(t+\Delta t_*) = v_{x}(t_{n_g}) +\frac{x\Delta v_R - y \Delta v_\varphi}{R}
\,,
\end{equation}

\begin{equation}\label{Eq-Vy_g_corr}
v_{y}(t+\Delta t_*) = v_{y}(t_{n_g}) + \frac{x\Delta v_\varphi + y \Delta v_R}{R}
\,,
\end{equation}

\begin{equation}\label{Eq-Vz_g_corr}
v_{z}(t+\Delta t_*) = v_{z}(t_{n_g}) +\tau [F_z(t+\Delta t_*) - F_z(t)]
\,,
\end{equation}
 where $v_x$, $v_y$, and $v_z$ are the components of the velocity vector of gaseous particles in the Cartesian coordinate system $(x,y,z)$,  $R=\sqrt{x^2+y^2}$, 
 $\Delta v_R = \tau [F_R(t+\Delta t_*) - F_R(t)]$, 
 $\Delta v_\varphi = \tau [F_\varphi(t+\Delta t_*) - F_\varphi(t)]$, 
 $\tau = 0.5 [\Delta t_* - \Delta t_g(t_{n_g})]$. 
 Here, $(F_R, F_\varphi , F_z)$ are the
components of the total gravitational force in the cylindrical system of coordinates $(R,\varphi , z)$.

For multiple GPUs, the details of a parallel OpenMP-CUDA implementation of SPH and N-body numerical algorithms are presented in the following papers~\citep{KhrapovKhoperskov2017SPH,Khrapov+2018Chel}.

\subsection{Stability Criteria}

Toomre~\cite{Toomre1964} derived a stability criterion that determines the growth of small-scale spiral perturbations. The criterion is:
\begin{equation}\label{eq:defQTstar}
Q_{T*} = \frac{c_r}{c_{T*}} \,, \quad c_{T*} = \frac{3.36 \,G \sigma_{*}  }{\varkappa} \,,
\end{equation}
where $\sigma_{*}$ is the surface density of stars, $\varkappa$ is the epicyclic frequency of the disk, and $c_r$ is the velocity dispersion of the stellar disk in the direction along the disk radius. For a gaseous disk, the criterion is:
\begin{equation}\label{eq:defQTgas}
Q_{Tg} = \frac{c_s}{c_{Tg}} \,, \quad c_{Tg} = \frac{\pi G \sigma_{g}  }{\varkappa} \,,
\end{equation}
where $c_s$ is the sound speed of gas and $\sigma_{g}$ is the surface density of the gaseous disk. Disks of real galaxies are multi-component, and the formulation of a local stability criterion for such systems is not an easy task. A few different local stability criteria have been suggested. We use the two-component criterion, as suggested by~\cite{RomeoFalstad2013Qstability}:
\begin{equation}\label{eq:defQTsum}
\Oo Q_{T\sum} = \frac{Q_{T*}Q_{Tg}}{Q_{Tg}+\mathcal{W}_g Q_{T*}} \,, \quad \mathcal{W}_g = \frac{2\,c_r\, c_s}{c_r^2 + c_s^2} \,.
\end{equation}
We note that gravitating disks are unstable with respect to large-scale spiral perturbations (global modes), even when the local $Q$-stability parameter is greater than unity, which, in particular, is the case  in our numerical simulations.

{The nonlinear stage of the large-scale instability that is discussed in Sections \ref{sec4.2} and \ref{sec4.3} re-distributes the surface density and velocity dispersion of the disk.}

\section{Simulations}
{It is convenient to choose the following units for the numerical modeling:}
\begin{equation}
	\begin{array}{c}
		\ell_m =  k_m \times 10^{10} M\ensuremath{_\odot} \,, \quad \ell_r =  k_r \times 10 \,\textrm{kpc} \,, \\
		\Oo \ell_v = \sqrt{G\frac{\ell_m}{\ell_r}} \simeq \sqrt{\frac{k_m}{k_r}} \times 65.75 \, \textrm{km s}^{-1}\,, \quad \ell_t 
\simeq \frac{k_r^{3/2}}{k_m^{1/2}}\times 142.7 \, \textrm{Myr}\,.
	\end{array}
\end{equation}

\textls[-5]{{With such a definition, the dimensionless gravitational constant and dimensionless mass of the stellar disk for any model will be equal to one for arbitrary values of the parameters $k_m$ and $k_r$ that were taken from the Table \ref{tab:ParametersModels}. 
We assume that the optical radius of the disk is equal to four radial scale lengths $R_{opt} = 4 r_d = \ell_r$, 
{where $r_d$ is the radial scale length of the thin stellar disk},
meaning that the dimensionless value of the optical radius of the disk is equal to the one as well ($\bar{R}_{opt} = 1$). 
For example, for models 706 and 707, we have $k_m = 3.72$, $k_r=0.9$ which gives $\ell_m = 3.72\times 10^{10} M\ensuremath{_\odot}$, $\ell_r = 9$\,kpc, $\ell_v \simeq 
133.7$\,km s$^{-1}$ and $\ell_t \simeq 63.2$\, Myr.}}

{Henceforth, we will use the dimensionless units unless the dimension of the quantity is explicitly indicated.}

\startlandscape
	\begin{specialtable}[H]
	\widetable
		\caption{Model parameters.}
		\label{tab:ParametersModels}
		\scalebox{0.80}{
		\begin{tabular}{ccccccccccccccccccc}
			\toprule
			\textbf{Experiment} & \boldmath{$M_d$ $\times 10^{10}$\,M$_\odot$}& \textbf{\boldmath{$r_d$} kpc}  & \boldmath{$\sigma_0$ M$_\odot/$pc$^2$}& \boldmath{$\sigma_\odot$ M$_\odot/$pc$^2$}& \textbf{\boldmath{$h$} pc}& \boldmath{$M_h$ $\times 10^{10}$\,M$_\odot$}& \textbf{\boldmath{$a$} kpc}& \boldmath{$\mu$ $\times 
10^{10}$\,M$_\odot$}& \boldmath{$M_b$} & \textbf{\boldmath{$b$} kpc}& \textbf{\boldmath{$r_b^{\max}$} kpc}& \boldmath{$M_g$ $\times 10^{10}$\,M$_\odot$}& 
\boldmath{$\mu_g$ M$_\odot/$pc$^2$}& \boldmath{$\sigma_{g\odot}$} & \boldmath{$\min{Q_{T*}}$}& \boldmath{$\min{Q_{Tg}}$} & \boldmath{$\min{Q_{T\sum}}$}\\\midrule
			\textit{\textbf{1}} & \textit{\textbf{2}} & \textit{\textbf{3}} & \textit{\textbf{4}} & \textit{\textbf{5}} & \textit{\textbf{6}} & \textit{\textbf{7}} & \textit{\textbf{8}} & \textit{\textbf{9}} & \textit{\textbf{10}} 
& \textit{\textbf{11}} & \textit{\textbf{12}} & \textit{\textbf{13}} & \textit{\textbf{14}} & \textit{\textbf{15}} & \textit{\textbf{16}} & \textit{\textbf{17}} & \textit{\textbf{18}} \\
			\midrule
			455 & 4.80 & 2.25 &1508  & 42.7 & 300 & 5.53 & 3.6 & 1.15 & 1.32 & 0.2 & 1.2 & 0.24& 0.05 &6.72  & 1.17/1.05 & 5.06 & 1.01\\ 
			470 & 3.61 & 2.25 & 1136 & 32   & 300 & 6.28 & 3.5 & 1.74 & 0.956& 0.16& 0.7 & 0.61 & 0.085 & 8.57 & 1.39/1.21 & 3.4& 1.16\\
			473 & 3.61 & 2.25 & 1136 & 32   & 300 & 6.28 & 3.5 & 1.74 & 0.956& 0.16& 0.7 & 0.61 & 0.085 & 8.57 & 1.25/1.12 & 3.39& 1.07\\
			474 & 3.61 & 2.25 & 1136 & 32   & 192 & 6.28 & 3.5 & 1.74 & 0.956& 0.16& 0.7 & 0.61 & 0.085 & 8.57 & 1.12/1.02 & 3.36& 
0.98\\
			475 & 3.61 & 2.25 & 1136 & 32   & 243 & 6.28 & 3.5 & 1.74 & 0.956& 0.16& 0.7 & 0.61 & 0.085 & 8.57 & 1.26/1.12 & 3.37& 
1.08\\
			481 & 4.87 & 2.25 & 1531 & 43.9 & 300 & 5.51 & 3.5 & 1.13 & 0.956& 0.16& 0.7 & 0.633 & 0.065 & 9.06 & 1.16/1.03& 3.81& 1\\
			482 & 4.87 & 2.25 & 1531 & 43.9 & 243 & 5.51 & 3.5 & 1.13 & 0.956& 0.16& 0.7 & 0.633 & 0.065 & 9.06 & 1.17/1.03 & 3.7& 1\\
			490 & 5.01 & 2.25 & 1576 & 44   & 300 & 5.23& 2.25& 1.04& 0.672& 0.16& 0.7 & 0.651 & 0.065&  9.07& 1.30/1.09 & 3.63& 1.06\\
			600 & 3.80 & 2.25 & 1194 & 33.7 & 300 & 6.03 & 3.8 & 1.59 & 0.964& 0.30& 1.4 & 0.642 & 0.085 & 9.02 & 1.36/1.22 & 3.3 & 
1.13\\
			601 & 3.80 & 2.25 & 1194 & 33.7 & 243 & 6.03 & 3.8 & 1.59 & 0.964& 0.30& 1.4 & 0.642 & 0.085 & 9.02 & 1.23/1.13 & 3.28 & 1.05\\
			602 & 3.80 & 2.25 & 1194 & 33.7 & 300 & 6.03 & 3.8 & 1.59 & 0.964& 0.30& 1.4 & 0.642 & 0.085 & 9.02 & 1.49/1.27 & 3.31 & 
1.21\\
			610 & 4.10 & 2.25 & 1290 & 36.8 & 300 & 5.97 & 2.81& 1.46 & 1.158& 0.30& 1.5 & 0.533 & 0.130 & 7.58 & 1.40/1.24 & 4.1 & 
1.19\\
			701 & 3.72 & 2.25 & 1170 & 33.4 & 315 & 5.90 & 3.33& 1.59 & 1.165& 0.30& 1.4 & 0.484&0.130 &13.7&1.27/1.14 & 2.18 & 1.05\\
			702 & 3.72 & 2.25 & 1170 & 33.4 & 315 & 5.90 & 3.33& 1.59 & 1.165& 0.30& 1.4 & 0.484&0.130 &13.7&0.98/0.95 & 2.18 & 0.88\\
			703 & 3.72 & 2.25 & 1170 & 33.4 & 388 & 5.90 & 3.33& 1.59 & 1.165& 0.30& 1.4 & 0.484&0.130 &13.7&1.32/1.19 & 2.19 & 1.09\\
			705 & 3.72 & 2.25 & 1170 & 33.4 & 388 & 5.90 & 3.0 & 1.59 & 1.159 & 0.31 & 1.6 & 0.484&0.130 &13.7 &1.34/1.21 &2.18  &1.11 
\\
			706 &3.72  &2.25 &1170 &33.4 & 388&6.02 &3.00 & 1.62 &1.01 &0.39 &1.8 &0.484 & 0.130&13.7 &1.33/1.2 & 2.19 & 1.10 \\
			707 &3.72  &2.25 &1170 &33.4 & 388&6.02 &3.00 & 1.62 &1.01 &0.39 &1.8 &0.391 & 0.105&11.1 &1.33/1.2 & 2.72 & 1.12 \\
			720 &2.72 &3.00 &481 &33.4 & 300&10.32 &3.01 & 3.79 &1.64 &0.21 &2.0 &0.412 & 0.151&13.7 &1.94/1.79 & 2.94 & 1.58 \\
			730 &3.08 &2.60 &725 &33.4 & 300&10.32 &3.01 & 3.79 &1.64 &0.21 &2.0 &0.505 & 0.164&13.7 & 1.59/1.44 & 2.58 & 1.32 \\
			\hline
		\end{tabular}
		}
		    \begin{tabular}{c}
\multicolumn{1}{p{\linewidth-1.5cm}}{\footnotesize \justifyorcenter{1---number of experiment; 2---the disk mass inside the optical radius ($r\le R_{opt}$); 3---the exponential scale of the stellar 
disk; 4---the central surface density of the stellar disk (See (\ref{eq:ExpSigmaStar})); 5---the surface density of the stellar disk 
near the solar neighborhood; 6---the vertical scale height (See (\ref{eq:VerticProfileDensity})); 7---the halo mass inside the 
optical radius; 8---the radial dark halo scale in the isothermal model; 9---the relative mass of the dark halo in units of the 
stellar disk within the optical radius; 10---the bulge mass; 11---the bulge scale length; 12---the bulge size; 13---the 
gaseous disk mass within $r\le 2R_{opt}$; 14---the relative mass of the gaseous disk in units of the stellar disk mass; 15---
the surface density of the gaseous disk near the solar neighborhood; 16---the minimum value of the Toomre parameter for the stellar disk; 
17---the minimum value of the Toomre parameter for the gaseous disk; 18---the minimum value of the Toomre parameter for the
two-component model.}}
\end{tabular}
	\end{specialtable}
	\noindent

\finishlandscape

When building equilibrium models of the Milky Way galaxy, it is important that the model parameters are consistent with
observational data. Below are listed the specific parameters and equilibrium profiles that must  be matched with the available observational data for the Milky Way.


\begin{enumerate}
	\item Rotation curve of the cold gaseous component $V_g(r)$.
	\item Rotation curve of the stellar disk $V_*(r)$.
	\item Velocity dispersion of the stars in the solar neighborhood $c_{r\odot}$ together with its radial dependence $c_r(r)$.
	\item Surface density of the thin stellar disk in the solar neighborhood $\sigma_{*\odot}$.
	\item Radial scale length of the density distribution of the thin stellar disk $r_d$.
	\item Surface density of the gaseous disk in the solar neighborhood $\sigma_{g\odot}$ together with the surface density dependence on radius $\sigma_{g}(r)$.
	\item Vertical scale height of the thin stellar disk $h$.
	\item Size of the stellar bar $r_{bar}$.
	\item Radial scale length of the bulge density distribution $b$ and its mass $M_b$.
\end{enumerate}

We have constructed over one hundred numerical models of the Milky Way galaxy using different sets of equilibrium parameters. Table \ref{tab:ParametersModels} lists the models satisfying the observational properties of the Milky Way. To discuss the results of the numerical simulations, we choose model 706 (see Table \ref{tab:ParametersModels}), which agrees with the available observational data for the Milky Way galaxy.

{Equilibrium particle distributions were constructed using the force balance equations. For the collisionless (stellar) particles, the Jeans equations were solved along the radial and vertical coordinates. For the gas particles, the hydrostatic-balance equations using a polytropic equation of state were solved in the $ z $ and $ r $ -directions. Because of the assumptions used to 
constrain the equilibrium state, and due to the finite number of particles modeling both the collisionless and the gaseous disks, the equilibrium solution is approximate and it contains so-called numerical noise. Thus, the evolution of a stellar-gas system begins from a quasistationary equilibrium that includes practically a full spectrum of initial perturbations. This instability leads to the growth of the most unstable modes that form the observed spiral structure.}

\subsection{Spiral Structure}

For the models considered, Figure~\ref{fig:Q_706_707} shows the typical radial dependence of the Toomre $Q$-parameter ($Q_{T}(r)$). There is a moderate increase of the $Q$-parameter in the region $r_d< r<4r_d$ with a broad minimum at the value $Q_{T\sum}=$\,(1--1.5). As is known from previous studies, in this case we have conditions for a spiral global instability and the formation of spiral structure in the disk, including a central bar.

Figure \ref{fig:706differTimes} shows the typical temporal evolution of a stellar gaseous disk with equilibrium parameters in agreement with observational data. The formation of an open spiral/bar in the central regions of the disk occurs rather quickly after approximately 1.5~disk revolutions and, after approximately two disk revolutions, the bar is formed. Outside the stellar bar, a complex spiral pattern grows. 
Even in the stellar disk, with its more regular spirals, a superposition of two- and three- armed spiral patterns with
different amplitudes grows and reaches the saturation stage. The gaseous disk exhibits an even more complicated spiral 
structure due to nonlinear interactions of the unstable modes, which leads to the branching of spirals, the appearance of rows, and that of a multi-level spiral pattern at the periphery of the disk.

{Fourier analysis of the surface density perturbations provides a means for visualizing the growing spiral and bar-like structures.}
It is convenient to characterize the growth of perturbations by the time dependence of the global Fourier amplitudes for
different azimuthal numbers $m$~\cite{Sellwood1986, Smirnov-Sotnikova2017}:

\begin{equation}
{A}_m(t)=\frac{1}{N_k} \sum\limits_{j=1}^{N_k} \exp\left\{ I\,  m\,\varphi_j(t)  \right\} \,,
\end{equation}
{where $I=\sqrt{-1}$, $\varphi_j$ is the azimuthal angle (radians) of the $j$-particle, and the summation is taken for the particles inside a radial layer $r_k \le r_j \le r_{k+1}$, $N_k$ is the total number of particles in this layer.}

Figures \ref{fig:sum_f_star} and \ref{fig:sum_f_gas} show the time dependence of the amplitudes of the $m =$\,(2--4) harmonics for different regions of the stellar and gaseous disks.
In the central regions of the stellar disk $r<2r_d$ 
{(see the left and middle panels in figure.~\ref{fig:sum_f_star})},
the dominant mode is $m=2$ harmonics, related to the bar. In addition to the $m=2$ mode, the disk is susceptible to the instability of $m=3$ and $m=4$ spirals, which are more noticeable at the periphery of the disk. The growth rates of perturbations in the central regions of the disk are noticeably larger when compared to that in the periphery of the disk. The temporal behavior of perturbations in the gaseous disk is qualitatively similar to that of the stellar disk. As one can see from Figure 6, in the central regions of the gaseous disk, the mode $m=2$ also prevails, while, outside the disk's central regions, the perturbations are represented by a combination of $m=2$, 
$m=3$, and $m=4$ spiral modes.

The number of spiral arms in the Milky Way galaxy, as well as their pitch angle, has been the subject of studies for a long time without consensus among the researchers as to the results. From observations of the stellar and the gaseous disks, 
estimates of the pitch angle of the Milky Way spiral pattern range from $5^\circ$ to $25^\circ$ 
\cite{Vallee2008SpiralArms,Vallee2017AngleSpiralArms,NikiforovVeselova2018SpiralStructure}.
Unsuccessful attempts to understand the properties of the global spiral pattern in the Milky Way disk have led to an approach where segments of the spiral arms are treated separately~\citep{Dame1986SpiralLMC,NikiforovVeselova2018SpiralStructure,XuEtAl2018SpiralGaiaMaser}. 
The disagreement between various determinations of the properties of the Milky Way spiral arms that are based on observational data (e.g., masers, ionized. neutral and molecular hydrogen gas, and 2MASS sources) that we suggest is a 
manifestation of the complex and non-stationary spiral structure of the Milky Way disk. We believe that this structure is a superposition of nonlinear spiral patterns with different azimuthal wavenumbers, angular speeds of the patterns, and amplitudes.
\end{paracol}
\nointerlineskip
\begin{figure}[H]
\widefigure{
	\includegraphics[width=0.7\textwidth]{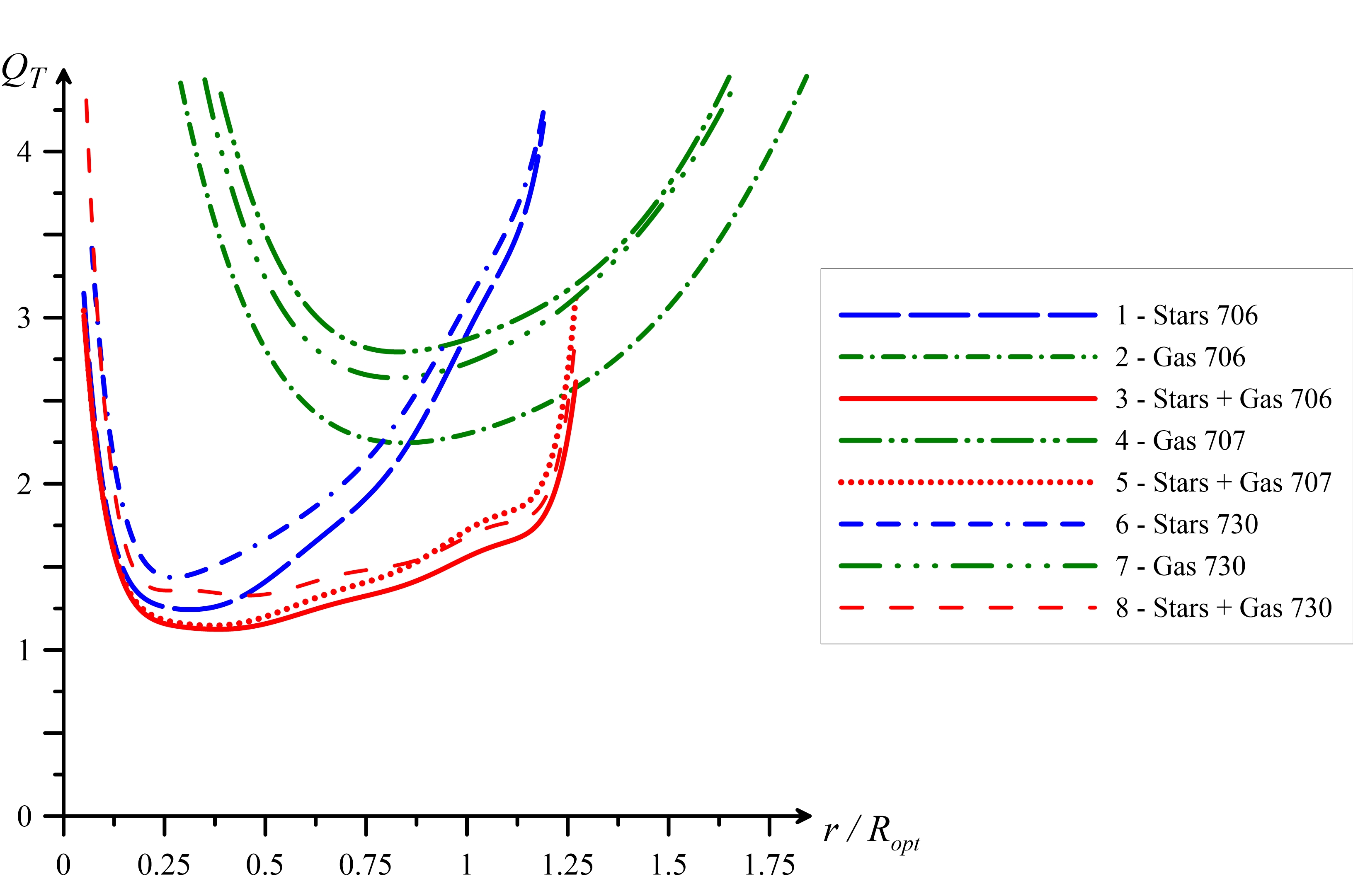} }
	\caption{
	{Radial dependence of the $Q_T$ parameters on the radius in two models. Curve $1$---$Q_{T*}$ parameter of the stellar disk (the same for both 706 and 707 models), curve $2$---$Q_{Tg}$ parameter for the gaseous component in model 706, curve $3$---two-fluid $Q_{T\sum}$ parameter for model 706, $4$---$Q_{Tg}$ parameter of gaseous disk in model 707, $5$---two-fluid $Q_{T\sum}$ parameter in model 707,
    curve $6$---$Q_{T*}$ parameter of the stellar disk in model 730, curve $7$---$Q_{Tg}$ parameter for the gaseous component in model 730, curve $8$---two-fluid $Q_{T\sum}$ parameter for model 730.}
}
	\label{fig:Q_706_707}
\end{figure}

\begin{figure}[H]
\widefigure{
	\includegraphics[width=1\textwidth]{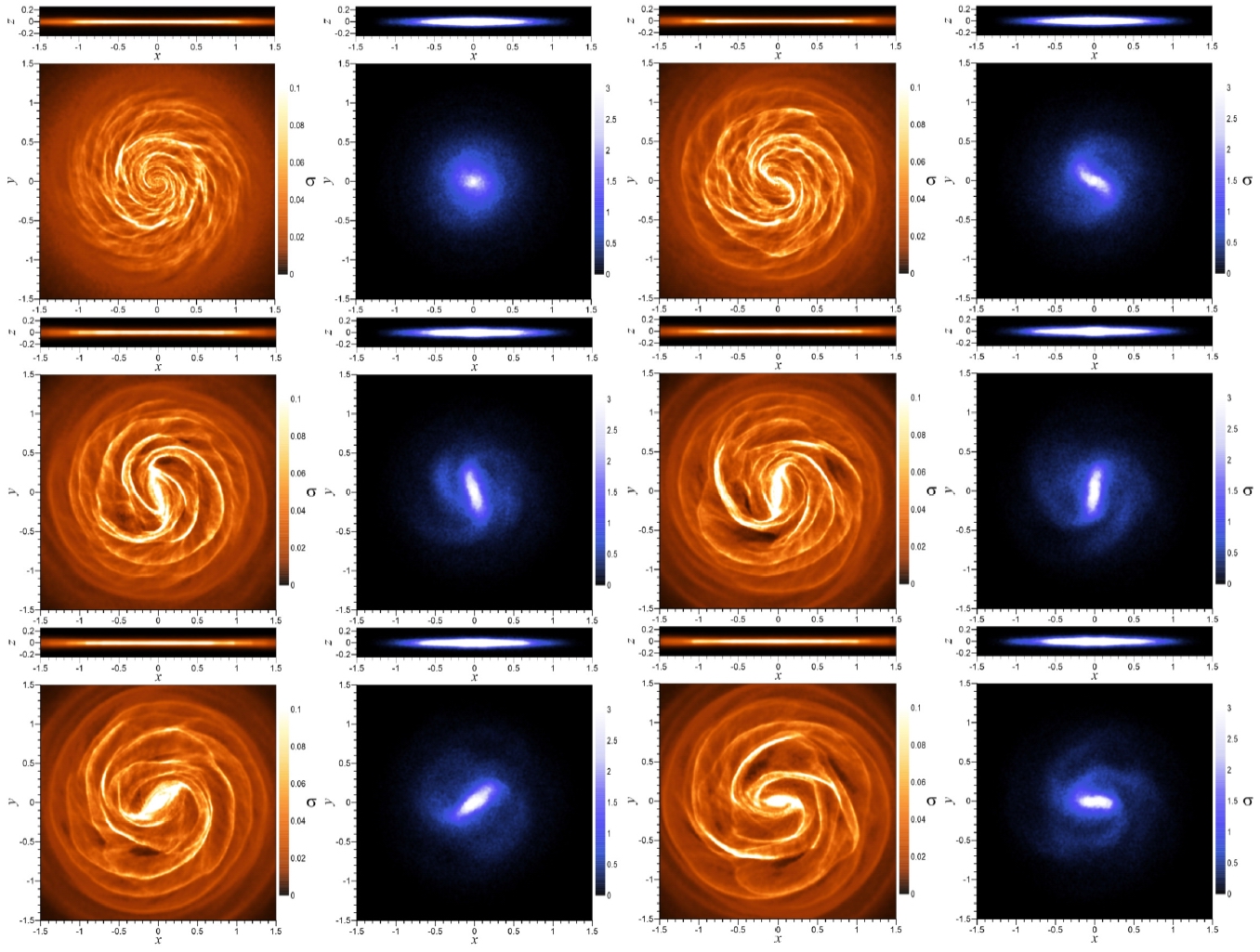} }
	\caption{Dynamics of gaseous
{(orange frames)} and stellar
{(blue frames)} disks at different times for model 706 shown sequentially, from left to right and from top to bottom, taken at times $t=\{3,7,11,15,19,23\}$. }
	\label{fig:706differTimes}
\end{figure}
\begin{figure}[H]
\widefigure
	\includegraphics[width=0.32\textwidth]{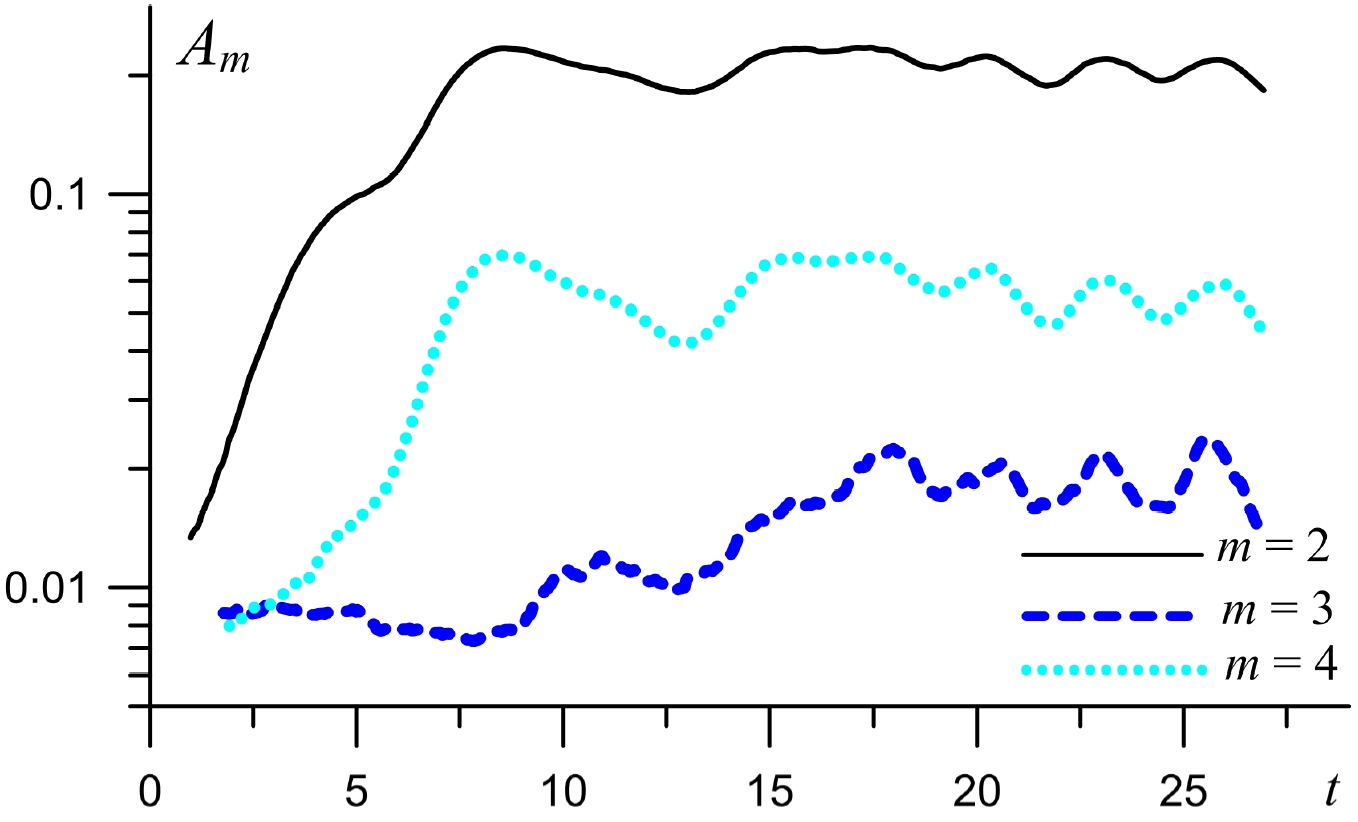}
	\includegraphics[width=0.32\textwidth]{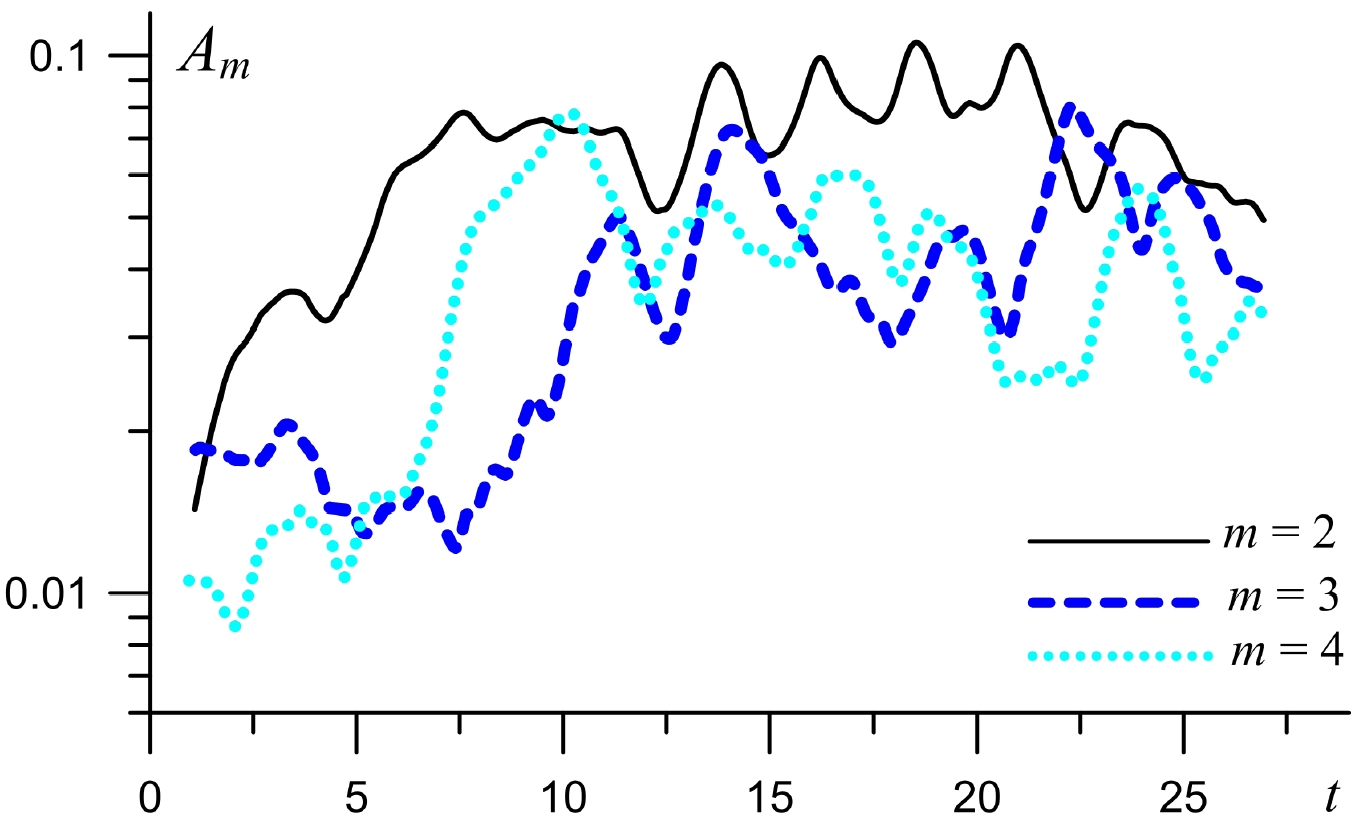}
	\includegraphics[width=0.32\textwidth]{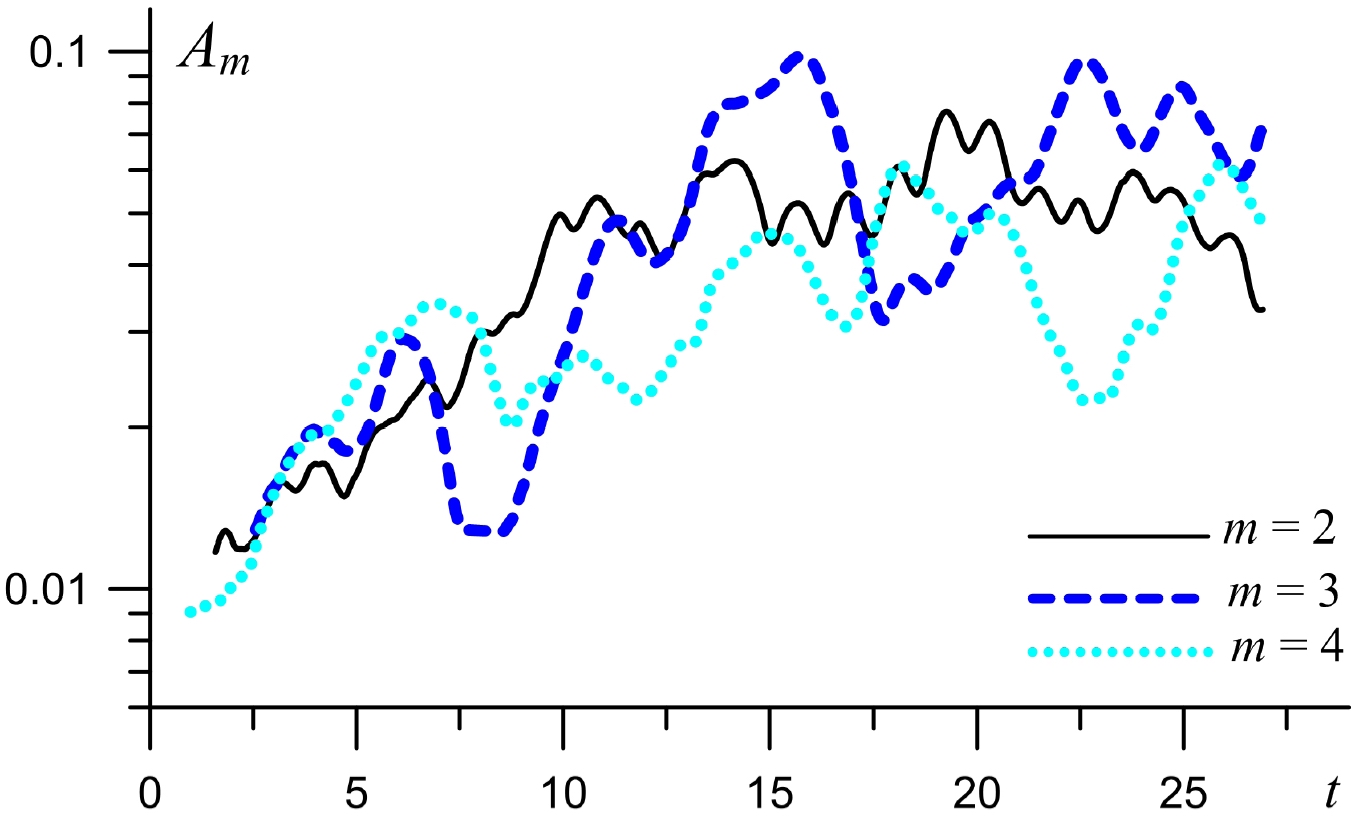}
	
	\caption{Time dependence of global amplitudes of $m=2,3,4$ Fourier harmonics in model 706 for the three regions of the stellar 
disk: (1) $r\le 1.8$\,kpc, (2) $3.6\le r\le 5.4$\,(kpc), (3) $7.2\le r\le 9$\,(kpc).}
	\label{fig:sum_f_star}
\end{figure}

\begin{figure}[H]
\widefigure
	\includegraphics[width=0.32\textwidth]{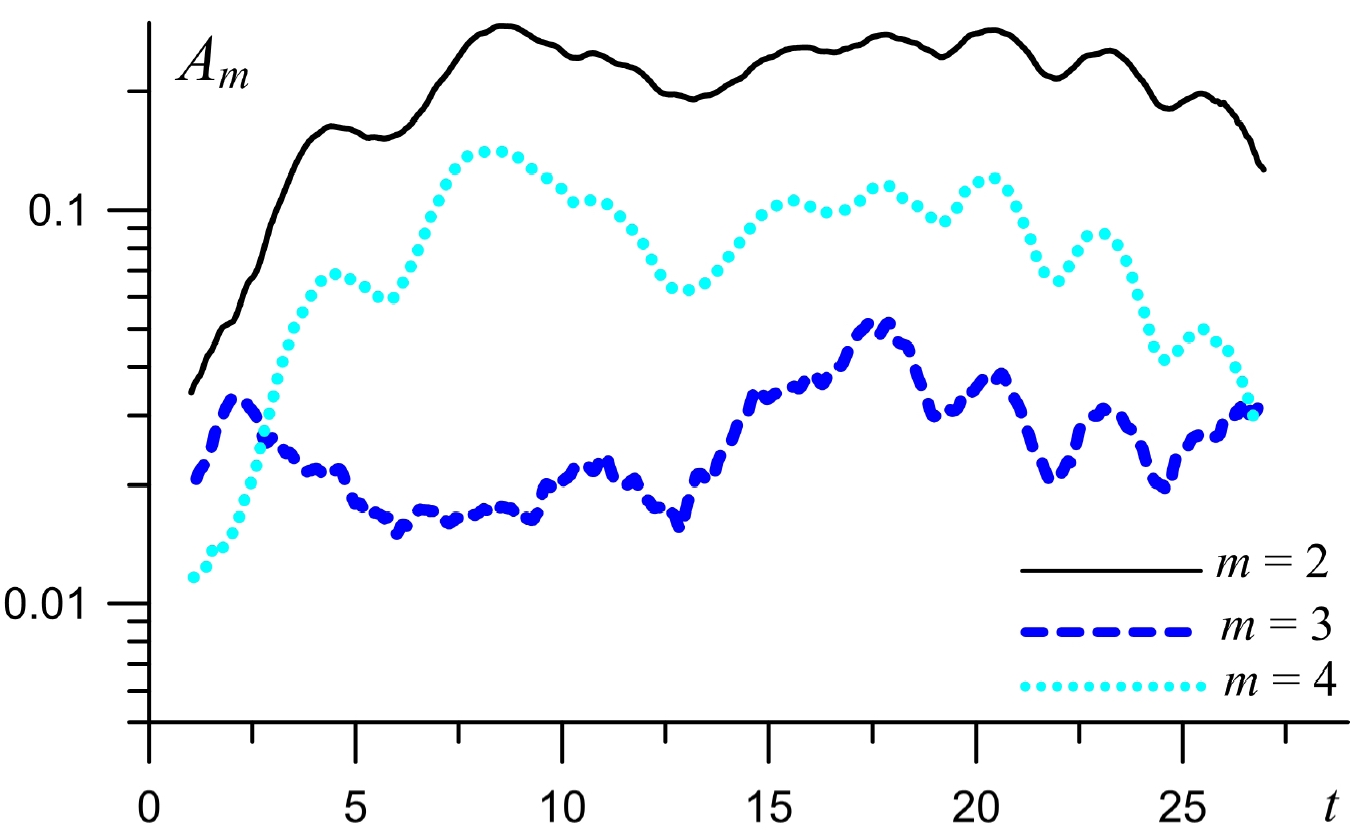}
	\includegraphics[width=0.32\textwidth]{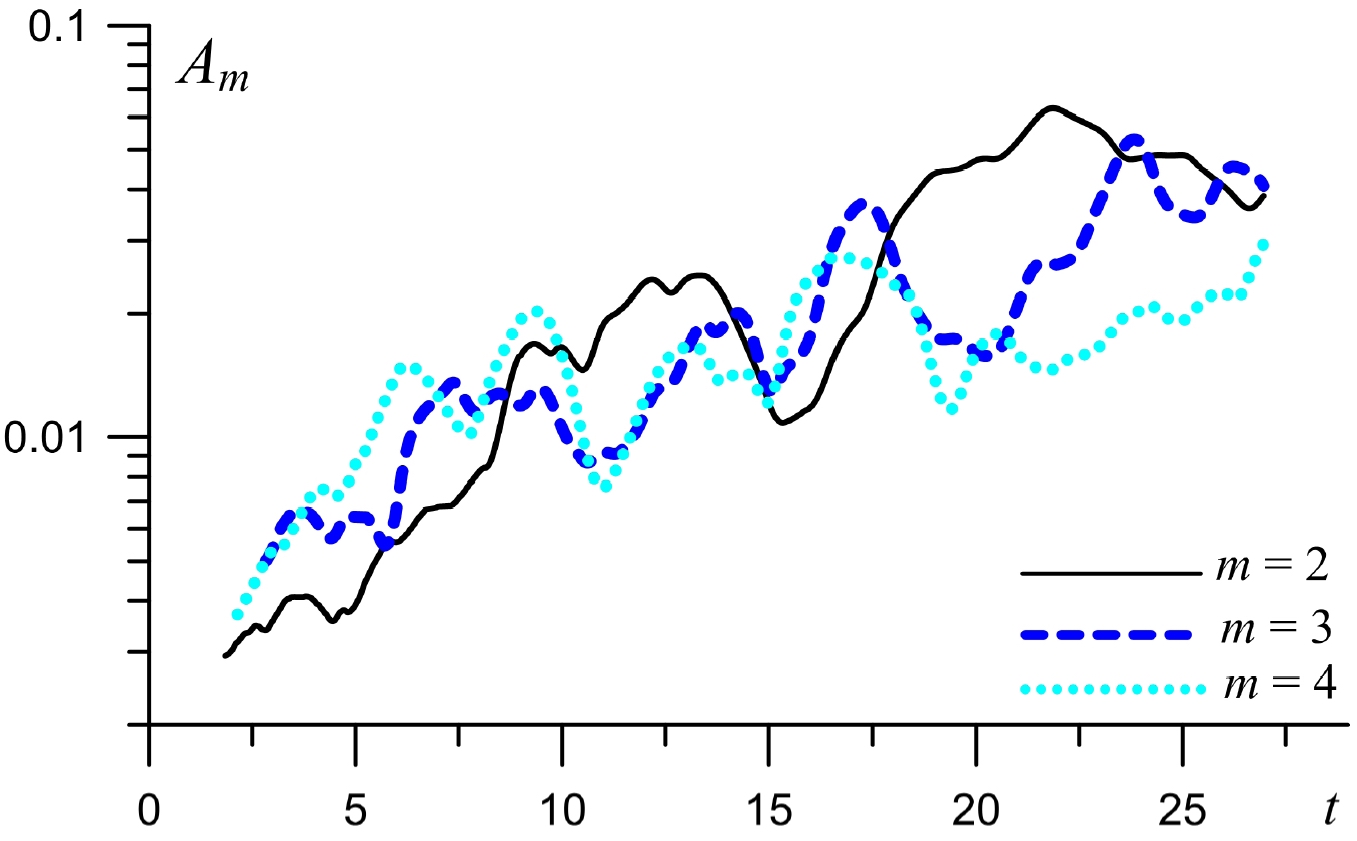}
	\includegraphics[width=0.32\textwidth]{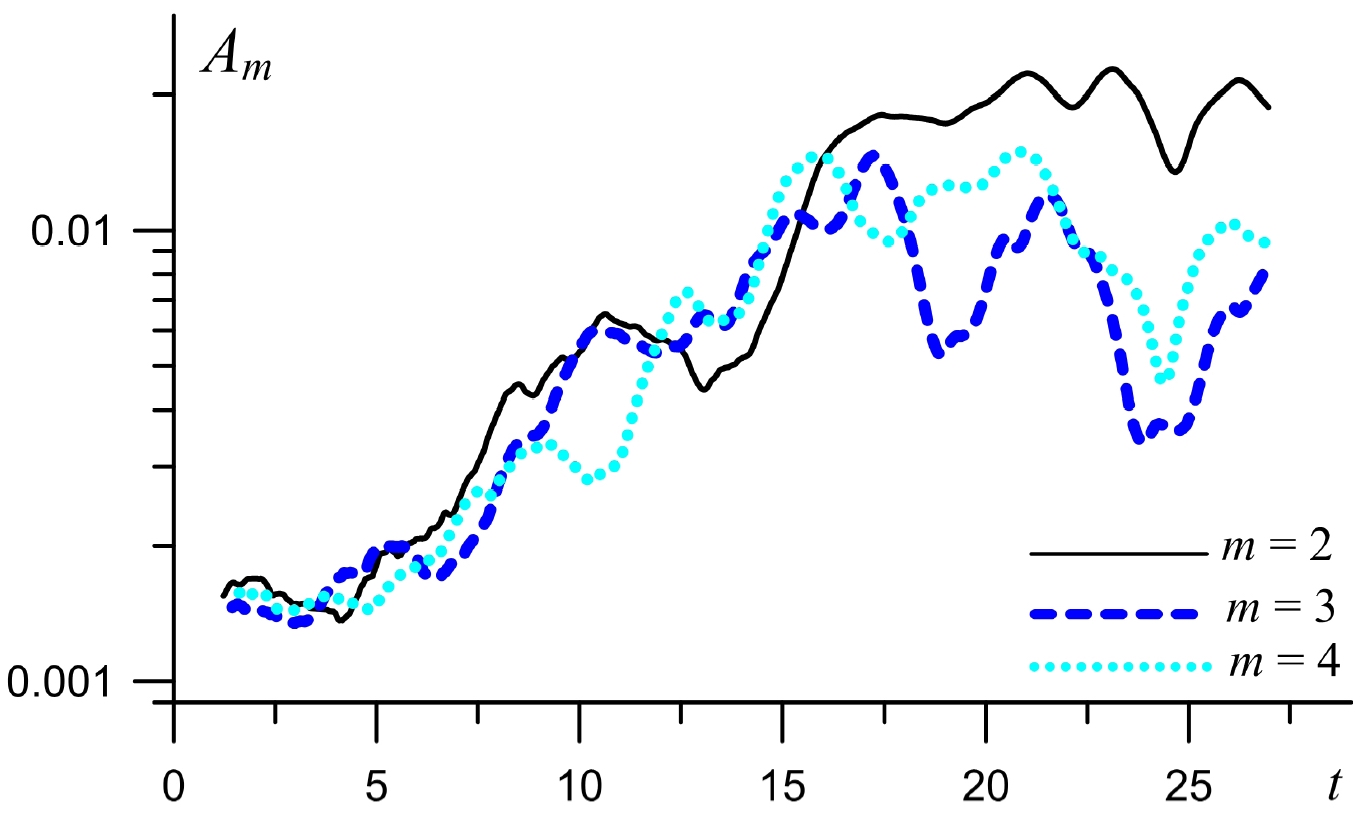}
	
	\caption{The same as in Figure~\ref{fig:sum_f_star}, but for the gaseous disk.}
	\label{fig:sum_f_gas}
\end{figure}

\begin{figure}[H]
\widefigure{
\includegraphics[width=0.99\textwidth]{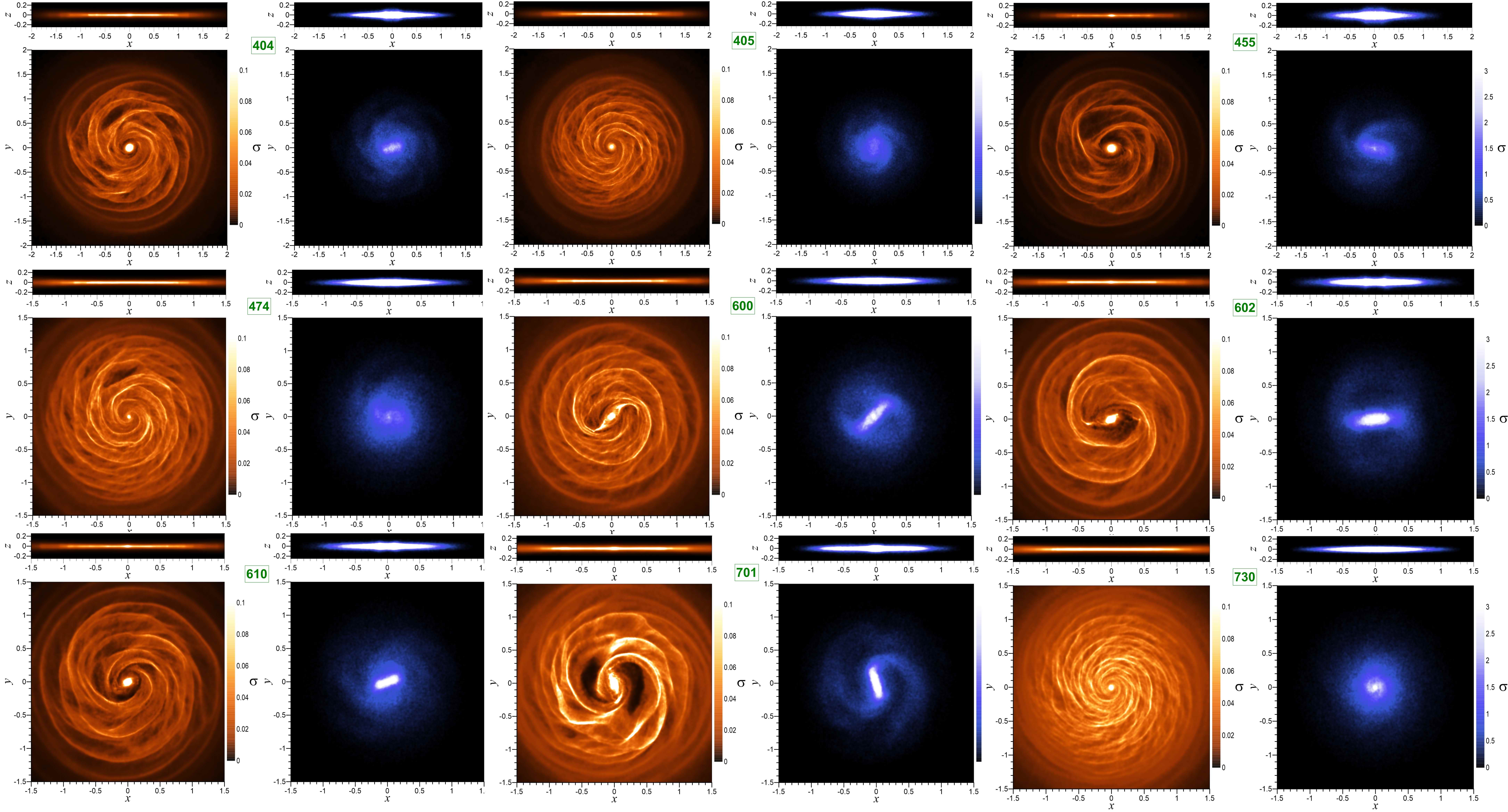} }
\caption{
{Morphology of the spiral structure of gaseous (orange frames) and stellar (blue frames) disks in different models (404, 405, 455, 474, 600, 602, 610, 701, and 730) at time $t=15$.}
}
	\label{fig:differSigmaModels}
\end{figure}

\begin{figure}[H]
\widefigure{
	\includegraphics[width=0.8\textwidth]{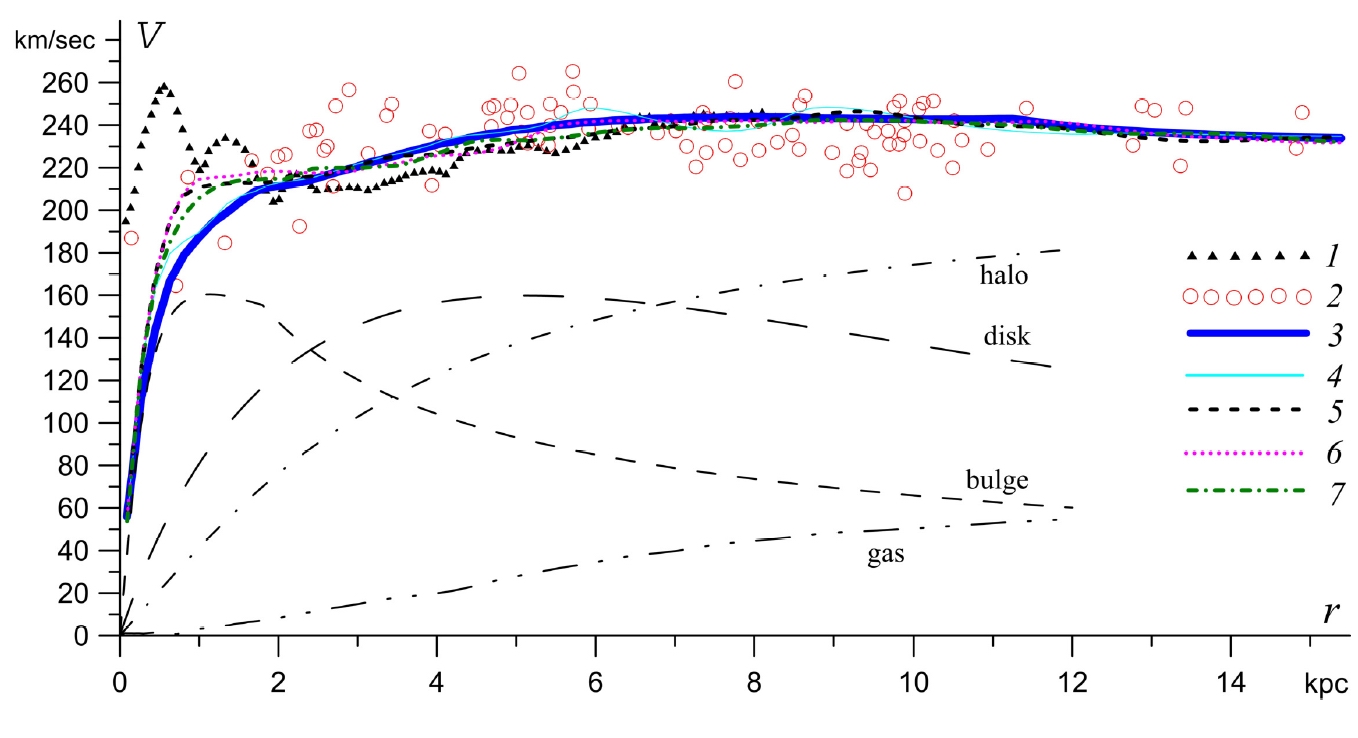} }
	\caption{Azimuthally averaged rotation curve of the gas in numerical model 706 at different moments of time compared to observational data: $1$---HI, $2$---masers, $3$---numerical model at $t=0$, $4$---$t=5$, $5$---$t=10$, $6$---$t=15$, $7$---$t=20$). }
	\label{fig:706kinem}
\end{figure}

\begin{paracol}{2}
\switchcolumn

Here we have paid special attention to the reconstruction of the  maximum of the Milky Way disk's rotational velocity in its central region $r\lesssim 1$\,kpc. Figure \ref{fig:differSigmaModels} illustrates the typical spatial distributions of the gas and of stars in models  401, 405, 455, and 474, where the rotational velocity in the disk has an internal maximum in the region $r\lesssim 1$\,kpc due to 
the bulge-disk mass distribution. We find that, due to a massive and centrally concentrated bulge, which is necessary for reconstructing the internal peak on the velocity rotation curve, the central bar either does not form or has a short major axis and is short-lived. Even if the bar appears at the initial stages of the disk evolution in a relatively cold stellar disk, it is destroyed at later stages.
A strong gradient in the gravitational potential in the central regions of the disk has a well-known 
\citep{Norman+1996BarDissolution,KatariaDas2018BarBulge} destructive role on bar formation that is caused by the presence of a compact massive bulge.
An additional factor suppressing bar instability is the increase of gas density in the central regions of the disk in the process of disk evolution. These arguments lead us to choose models in which the central peak of the disk's rotation curve is not explained by the presence in the disk of a massive and compact bulge. Instead, the central peak of the disk's rotation curve is formed by the process of nonlinear evolution of spiral/bar perturbations. 
Examples of these models are  600, 602, 610, 701 shown in Figure \ref{fig:differSigmaModels} and 706 in Figure~\ref{fig:706differTimes}.
These models develop a pronounced stellar bar with gas kinematics that are similar to that observed for the Milky Way bar.

Outside the central regions of the stellar disk, models with a bar demonstrate a mixture of two-, three-, and four-armed spiral patterns  of lower amplitude. The gaseous disks show a more complicated behavior, with the appearance of a non-stationary spiral pattern.

We note that models with a relatively low stellar velocity dispersion (e.g., model 701 in Table \ref{tab:ParametersModels}) develop unrealistically high-amplitude spiral perturbations. Actually, the development of a strong spiral perturbations leads to a heating of the stellar disk, and the velocity dispersion of the stars increases to approximately 40\,km\,s$^{-1}$ (see {Figure~\ref{fig:cr_time})} in a short period 
of time. Thus, the observed amplitudes of spiral perturbations in the disk of the Milky Way impose additional restrictions on the velocity dispersion of stars in the solar neighborhood. Models with a velocity dispersion $c_{r\odot}=$\,(29--33)\,km\,s$^{-1}$ 
support a long-lived stellar bar with a semi-axis of 
$\sim$3 kpc, together with a spiral pattern that has observable amplitudes. 
Edge-on images of the numerical models demonstrate features that are similar to those observed in the central regions of the Milky Way, such as the so-called X-structure (see the edge-on images of models 401, 455, 602, 610, and 701 in Figure~\ref{fig:differSigmaModels}.

The radial scale length of the stellar disk is of critical importance in the development of a bar in the disk's central regions. 
The models with a radial scale length of the stellar disk larger than $r_d=2.6$\,kpc, such as model 730 with $r_d=2.6$\,kpc or model 720 with $r_d=3$\,kpc, lead to a relatively low surface density disk in the central region $\sigma_{0}\lesssim 700$\,$M_\odot$\,pc$^{-2}$. 
That, in turn, requires the presence of a massive dark matter halo to explain the observed disk rotation curve. Such models do not allow the formation of a bar-mode in the disk
(see, e.g., model 730 in Figure~\ref{fig:differSigmaModels}).
{Besides that, for model 730, the values of the parameters $Q_{T*}$ and $Q_{Tg}$ are (15--20)\% larger within the optical radius when compared to models 706 and 707, which also leads to the stabilisation of the bar-mode.}
Attempts to explain the observed central maximum of the rotational velocity of the disk in these modes (curves  \textit{1}, \textit{2} in Figure~\ref{fig:RotObs}) were unsuccessful. We then considered a set of models with a less concentrated bulge, which explained the appearance of the maximum of the disk's rotational velocity as a result of bar appearance and the nonlinear evolution of perturbations.

\end{paracol}
\nointerlineskip
\begin{paracol}{2}
\switchcolumn
\vspace{-8pt}

\subsection{Density Profile and Rotation Curve}\label{sec4.2}

Here, we discuss the behavior of model 707 (see Table \ref{tab:ParametersModels}), which satisfies all of the limitations imposed by the observational data for the Milky Way galaxy, as discussed above.
Figure~\ref{fig:706kinem} shows the decomposition of the rotation curve for this model.

As one can see from Figure \ref{fig:706kinem}, the nonlinear stage of instability leads to a steepening of the gas's rotation curve in the disk's central regions. However, note that the amplitude of the central peak in the gaseous rotation curve is smaller than that observed for the velocity peak in the rotation curve of the Milky Way gaseous disk.

{Figure \ref{fig:706azimutDensStarGas} shows the azimuthal density profiles of the stellar  $\sigma_*(\varphi)$ and of the gaseous disks $\sigma_g(\varphi)$ at three fixed radii
$r=0.15$ (inside the bar), $r=0.51$ (at two exponential scale lengths of the disk), and at the periphery of the stellar disk close to the Sun's orbit $r=0.89$. Dashed black lines correspond to the averaged values of the density over the azimuthal angle $\varphi$. Azimuthal density distributions demonstrate the features of the spiral disturbances. In the central zone, the stellar density distribution shows a pronounced two-arm mode. As for the gaseous component, in addition to the main mode, it shows the presence of two other 
modes at lower amplitudes.
At two radial scale lengths of the disk ($r \simeq 2r_d$), the stellar system demonstrates a four-armed spiral structure.
At the periphery of the stellar disk, the amplitudes of the density perturbations decrease considerably and remain within 10--20 percent.
Gas shows more complicated behavior, which can be described as a superposition of two-, three-, and four-armed spiral patterns (see Figure~\ref{fig:sum_f_gas}).
 }

The radial scale length of the stellar density distribution is one of the key parameters that determines the mass of the stellar disk and its stability properties. The estimated radial scale length of the Milky Way's stellar density distribution is about 2.25 kpc 
\cite{DrimmelSpergel2001MW} and we use that value in our simulations. In the set of models with $r_d=3$\,kpc and  $r_d=2.6$\,kpc, the surface density of the stellar disk decreases in the central regions to $\sigma_0=481\,M_\odot/$pc$^2$ and $\sigma_0=725\,M_\odot/$pc$^2$ (corresponding to models 720 and 730). In order to satisfy the observed rotation curves in these models, one has to increase the mass of the dark matter halo relative to the mass of the disk to $\mu\simeq 5$. This, as a result, suppresses the formation of the central bar in the Milky Way galaxy. In other words, with the observed surface density of the solar neighborhood, the dynamical models do not allow us to reproduce the stellar bar in the central regions of the disk if the exponential scale length of the surface density distribution is more than 2.5~kpc.

We find that the dynamical models that conform not only to the observed rotation curve of the galactic disk, but also to the 
surface densities of the stellar and gaseous components in the solar neighborhood, together with their observed velocity 
dispersions, give a relatively small total mass of the disk $M_{d}=M_{*}+M_{g}\simeq 4.5\times 10^{10}\, M_\odot$ within 
$2R_{opt}=18$\,kpc.
Therefore, dark matter dominates within $2R_{opt}=18$\,kpc, composing about 74 percent of the total mass of the Milky Way galaxy.
\end{paracol}
\nointerlineskip
\unskip
\begin{figure}[H]
\widefigure{
	\includegraphics[width=0.85\textwidth]{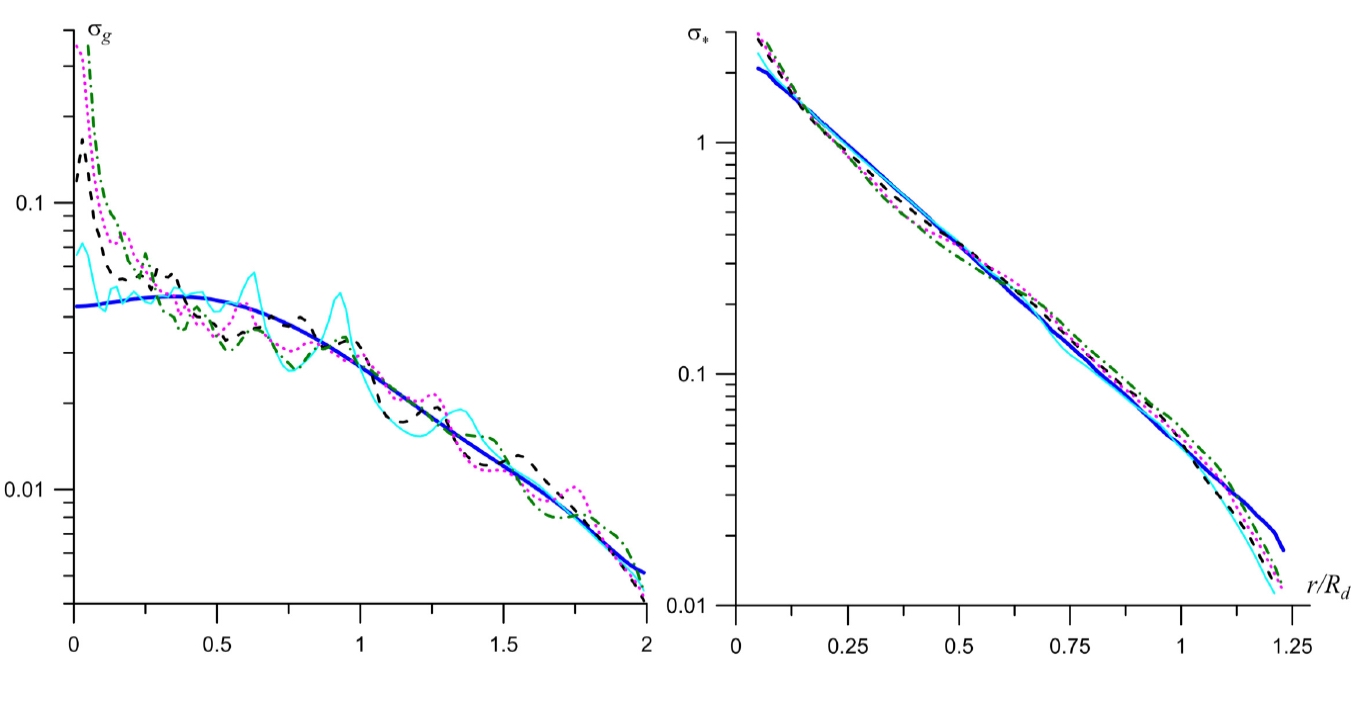} }
	\caption{{Azimuthally averaged} radial dependencies of the surface densities of gas $\sigma_{g}$ and stars $\sigma_{*}$ at different times. The times shown are the same as in Figure \ref{fig:706kinem}.
  }
	\label{fig:706density}
\end{figure}

\begin{paracol}{2}
\switchcolumn

\begin{figure}[H]
{
	\includegraphics[width=0.51\textwidth]{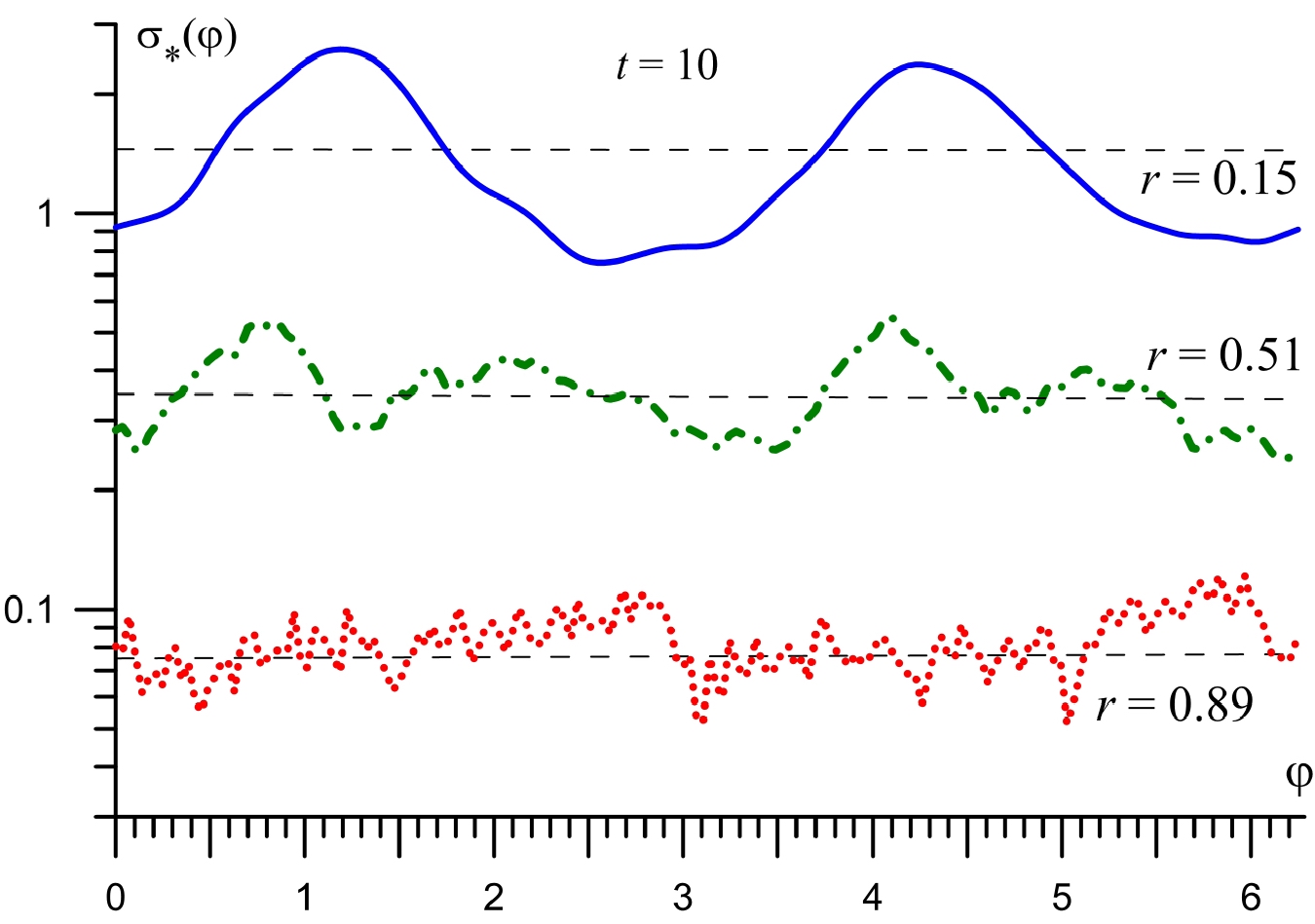}
	}
	{
	\includegraphics[width=0.51\textwidth]{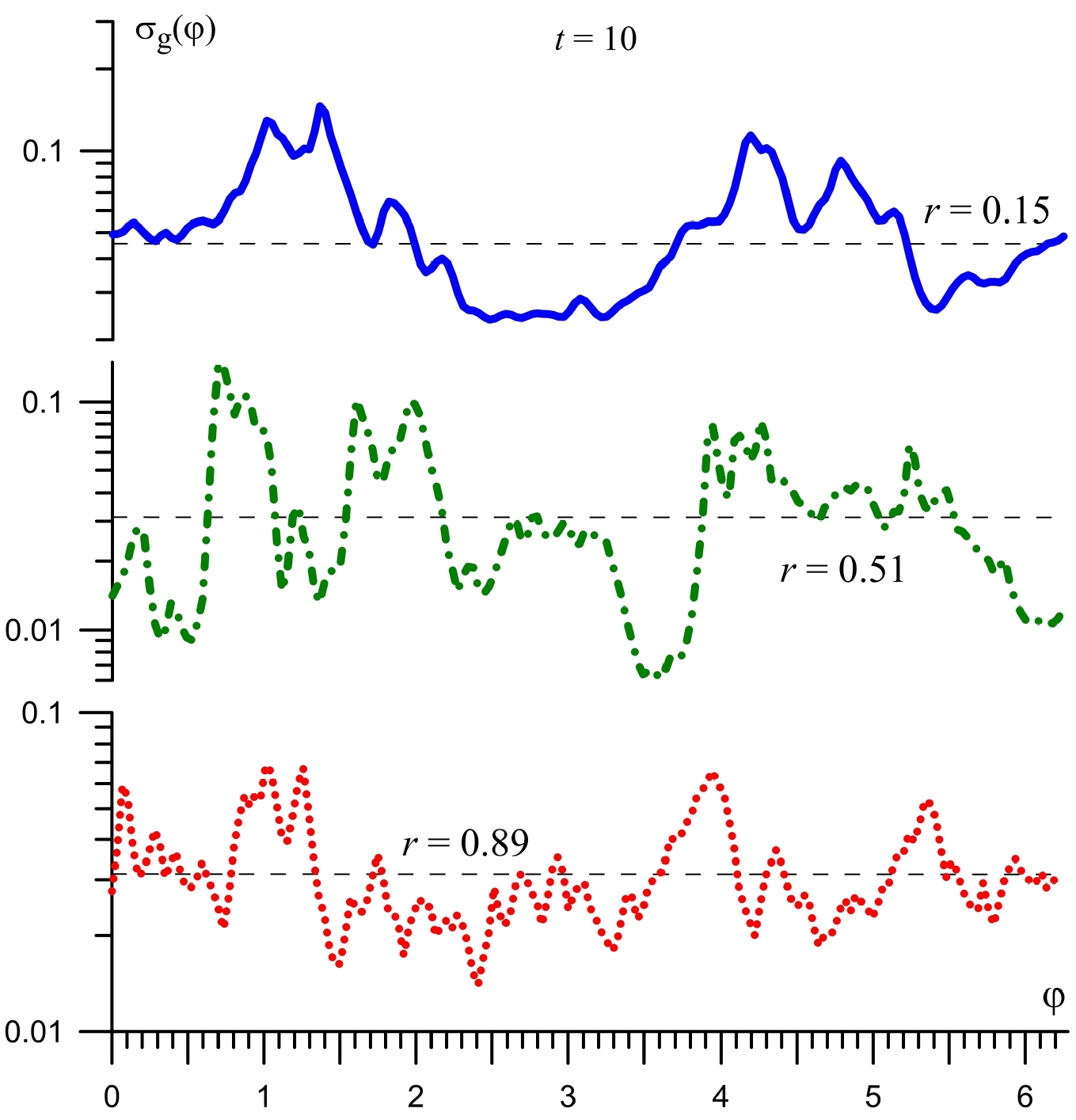}
	}
	\caption{
	{Azimuthal dependences of the logarithm of the density distributions for stellar (upper panel) and for gaseous (lower panel) components taken at three different radii at time $t=10$ (experiment 706). The azimuthal angle   $\varphi$ is in radians}.
}
	\label{fig:706azimutDensStarGas}
\end{figure}


\subsection{Stellar Velocity Dispersion}\label{sec4.3}

As an initial condition, we choose a radial dependence of the velocity dispersion that corresponds to the observational data for the Milky Way (curve 6 on Figure~\ref{fig:DispObs}). At the galactocentric radius of the Sun, this choice gives the value
$c_{r\odot} \simeq$\,(30--32)\,km\,s$^{-1}$.
The
{tangential}
component of the velocity dispersion in the collisionless disk $c_\varphi$ is prescribed according to Equation (\ref{eq:crcfKappa}).

During the evolution of the perturbations, the velocity dispersion of the stars evolves. Figure~\ref{fig:cr_time} illustrates the temporal evolution of the velocity dispersion for a number of models. As one can see, the typical behavior of the velocity dispersion is its slow evolution during approximately one billion years. At roughly this time, 
the formation of spiral structure and of a bar takes place, and it is accompanied by essentially no heating of the collisionless disk. At later times, some increase of the velocity dispersion of the disk is~observed.

\begin{figure}[H]
{
	\includegraphics[width=0.5\textwidth]{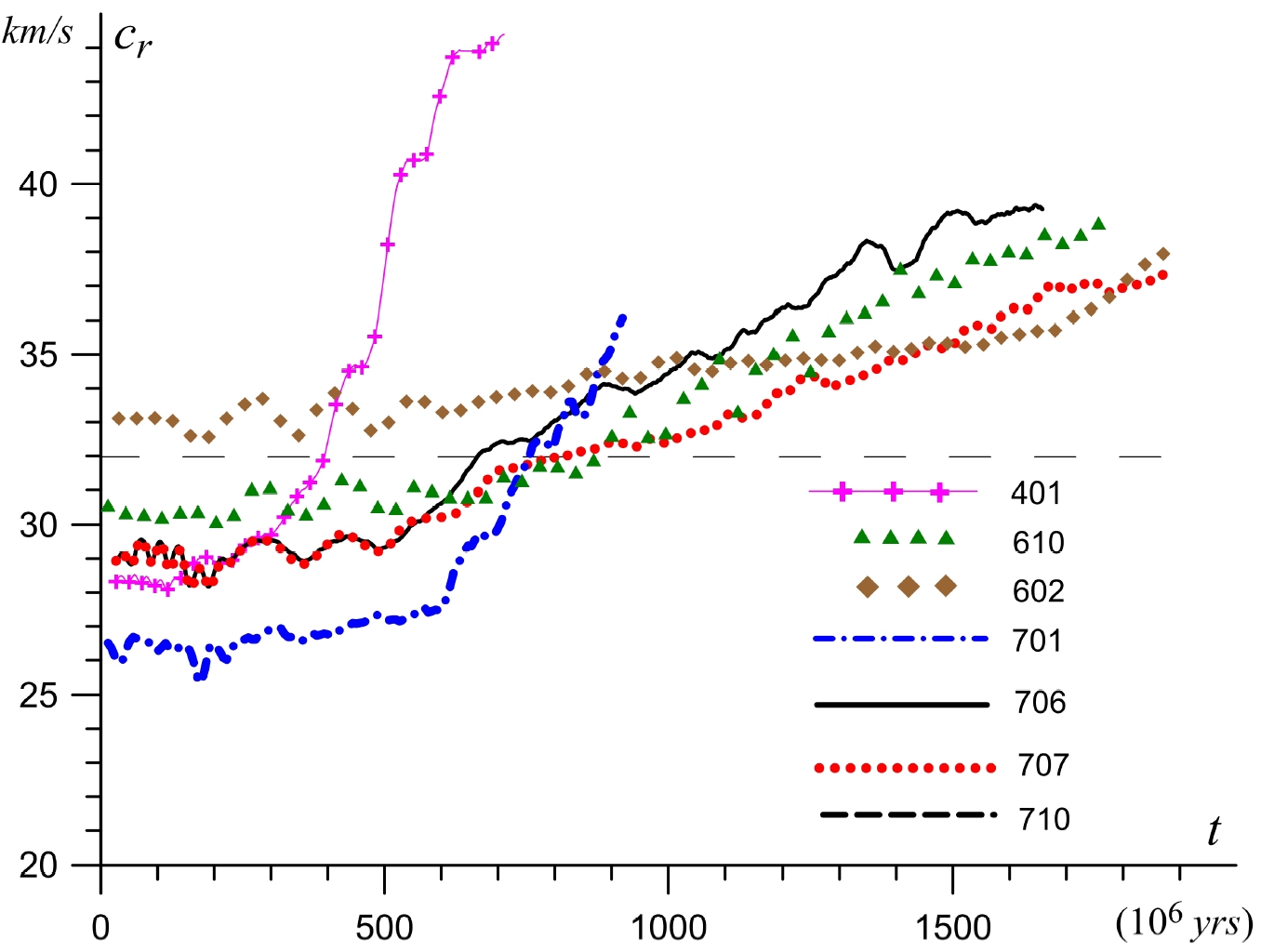} }
	\caption{Dependence on time of the velocity dispersion $c_{r\odot}(t)$ for different models. 	}
	\label{fig:cr_time}
\end{figure}

During the development of perturbations, the velocity dispersions deviate from those of the axisymmetric case. 
Figure~\ref{fig:cr_phi} shows the dependence of the velocity dispersion $c_r$ as a function of the azimuthal angle $\varphi$ at three different radii of the disk for model 707 at time $t = 9$. As one sees, large amplitudes of the density perturbations in the central regions of the disk are accompanied by strong variations in the velocity dispersion $c_r$. The variations of the velocity dispersion are less prominent at the periphery of the disk.

\begin{figure}[H]
{
	\includegraphics[width=0.5\textwidth]{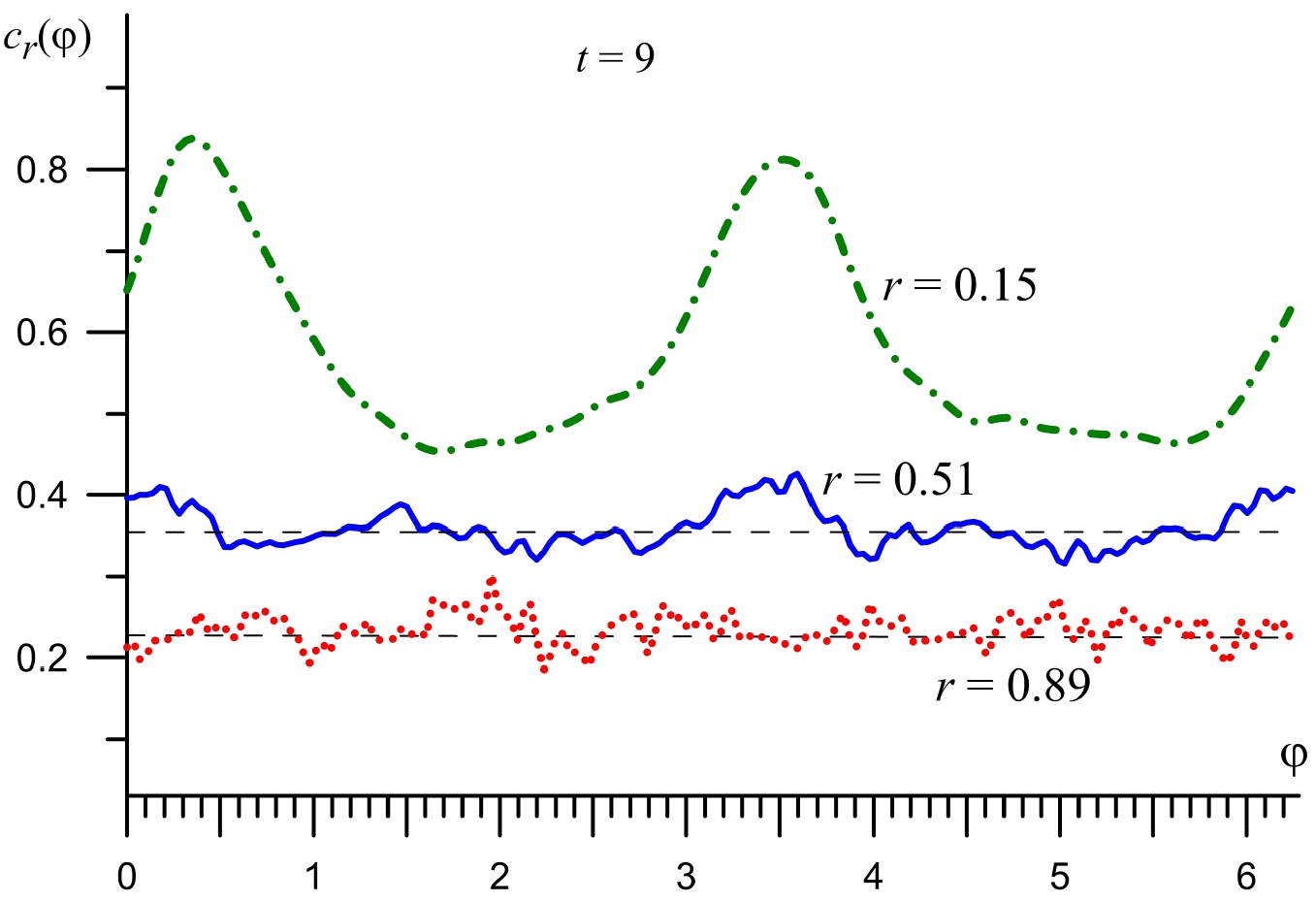} }
	\caption{Azimuthal dependencies of the velocity dispersion $c_{r\odot}(\varphi)$ at different radii for model 707.	
	{The azimuthat angle $\varphi$ is measured in radians}.
	}
	\label{fig:cr_phi}
\end{figure}

\subsection{Disk Kinematics in Central Regions}

Because of presence of the bar, the kinematical properties of the stellar and gaseous disks inside $r\lesssim 3$ kpc demonstrate 
some specific features. Figure~\ref{fig:Vstar_bar} shows the rotation curves of the disk projected along lines representing different viewing orientations with respect to the semimajor axis of the bar. As one can see, the rotation curve of the disk along the semi-minor-axis of the bar has a local maximum inside 1 kpc from the center of galaxy, similar to the observed kinematical properties of the Milky Way disk within the central kiloparsec (see Figure~\ref{fig:RotObs}).

The estimates of the angular velocity and mass of the Milky Way bar~\citep{Michtchenko2018MWfullModelbar} provide values of
$\Omega_{bar}<50\,$km\,s$^{-1}$\,kpc$^{-1}$ and $M_{bar}\simeq 2\times 10^9\,M_\odot$, respectively.

The conditions for the formation of a bar are rather sensitive to the mass of the dark matter halo $\mu$ and that
of the bulge $\mu_b$, relative to the mass of the thin disk. An important parameter is the radial scale 
length of the disk density distribution together with the scale length of the mass distribution of the bulge, as  mentioned above. In the models with $\mu \gtrsim 2$, the bar mode does not develop due to the suppression of bar instability by the massive halo. 
{The classical stability criterion of the global bar mode 
\cite{Ostriker-Peebles-1973bar} 
requires that the ratio of the kinetic energy of disk rotation $E_{rot}$ to the total gravitational energy $|E_{\Phi}|$ does not exceed some critical value $(E_{rot}/|E_{\Phi}|)_{crit}\simeq 0.14$. Thus, galaxies with a dark halo are more stable with respect to the formation of the bar mode. 
The authors~\cite{Ostriker-Peebles-1973bar}
 made a fundamental conclusion, that stabilization of the bar mode requires the value of the parameter $\mu$ within (1--2). Criteria of this kind have been tested using better N-body models and they show that the exact value of  $\mu$ depends on a number of parameters of the galactic subsystems that determine the properties of the bulge, the gaseous disk, the radial profile of the dark matter, etc.~\cite{Polyachenko2016, zasov-etal2017, Michtchenko2018MWfullModelbar, Shen-Zheng-2020structureMW, sak-2020}. 
 }

\end{paracol}
\nointerlineskip
\begin{figure}[H]
\widefigure
	\includegraphics[width=0.49\textwidth]{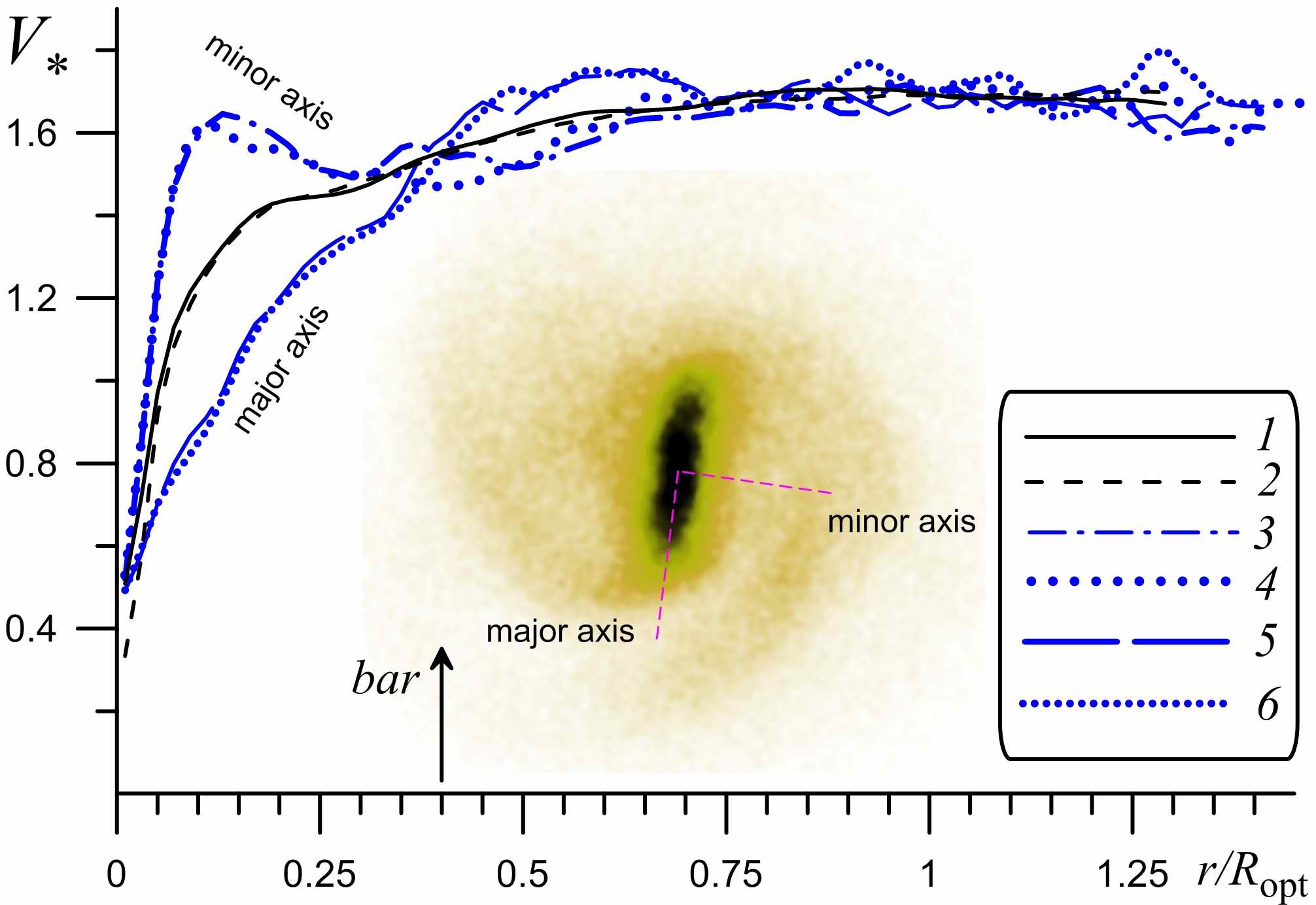}
	\includegraphics[width=0.49\textwidth]{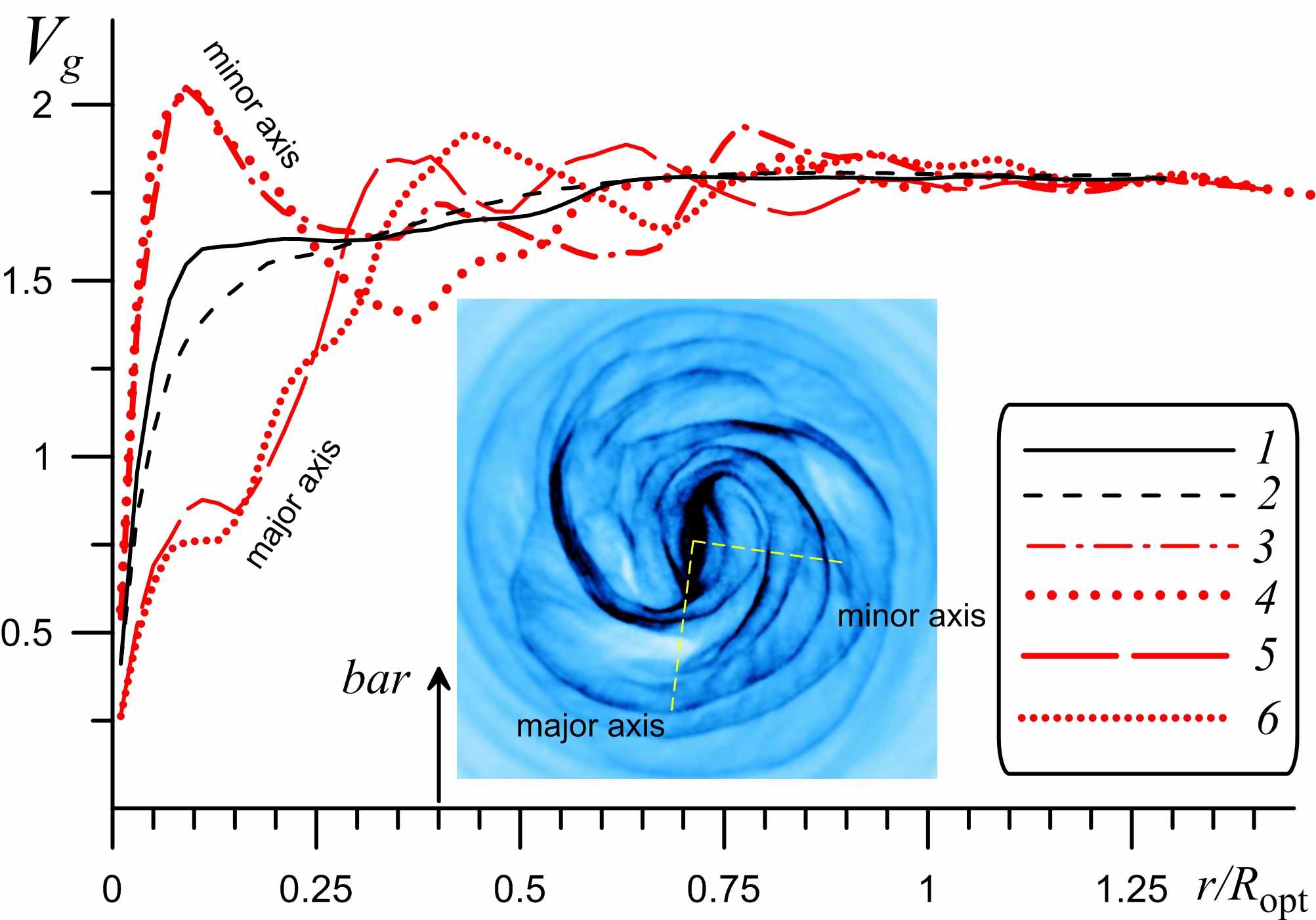}

	\caption{The typical profiles of the azimuthal velocities of the stellar
{(left frame)} and gaseous disks
{(right frame)} along the minor (curves 3 and 4) and major (curves 5 and 6) axis of the bar, taken at $t=15$ for model 
706.
{We show two rotation curves along major and minor semi-axes of the bar.}
For comparison, the azimuthally averaged velocity profiles are also shown. Curve 1 shows the azimuthal velocity at $t=0$, while $2$ is the azimuthally averaged rotational curve at time $t=15$. The vertical arrow shows the approximate size of the bar.}
	\label{fig:Vstar_bar}
\end{figure}
\begin{paracol}{2}
\switchcolumn

{Let us estimate the influence of the relative mass of the halo on the velocity dispersion of stars that are close to the boundary of stability of the system. The rotational velocity of the the disk is
$\displaystyle V \simeq \sqrt{R\frac{\partial \Phi}{\partial R}}$ where $R$ is some radius, and $\Phi$ is the gravitational potential. Distracting from the particular mass distribution, we can write:
${\partial \Phi}/{\partial R}\sim G (M_d + M_h)/R^2$.
On the boundary of stability, the velocity dispersion of stars is inversely proportional to the epicyclic frequency: 
$c_r \propto 1/\varkappa\propto 1/\sqrt{M_d + M_h}$. As a result $c_r/V \propto 1/(M_d+M_h) \propto 1/(1+\mu)$ for the fixed mass of the disk $M_d$, so the disk at the boundary of stability is colder in the models with a more massive halo.
 }

Similarly, in the models with a moderate ratio of masses of the halo and of the disk $\mu\simeq 1.6$, but with a massive and compact bulge $b/R_{opt} \gtrsim 0.014$, the bar is not reproduced in the numerical simulations due to scattering of the orbits of the collisionless particles in the central regions where the gravitational potential has a large gradient. Thus, the presence of a bar in the Milky Way disk puts rather strict limitations on its equilibrium properties.

\section{Discussion}

{Athanassola~\cite{Athanassoula2002} showed that a bar grows faster in galactic models with a live halo as compared to models with a rigid one. Athanassoula attributed this effect to the destabilizing influence of the resonant halo particles. The total number of particles in Athanassoula’s numerical simulations was of order of $10^6$ with the mass of individual particle being about $2.5\times 10^5$ M$_\odot$.  This leads, apart from the resonant effects, to the generation of high amplitude noise in the disk component. Polyachenko et al.~\cite{Polyachenko2016} returned to the problem of bar formation using one to two orders of magnitudes larger numbers of particles. They found that, while a live halo decreases the time of growth for the bar instability from 500 to 250 Myr, the pattern speed and other parameters of the bar remain approximately the same.  Because we are interested in the late stages of disk evolution, the possible influence of a live halo on the growth rate of the bar instability is insignificant for our purpose. 

{A possible influence of the "live" halo on the dynamics of the disks can be obscured in the numerical simulations by the particle resolution when massive dark matter "particles" generate a high amplitude noise in the disk. }

}

{Long et al.~\cite{Long2014} and Collier et al.~\cite{Collier2019}  showed that spinning dark live halos suppresses the growth of stellar bars. Recent studies based on GAIA DR2 observational data~\cite{Bajkova2020} demonstrate that the Milky Way halo has close to zero rotational velocity with average value of $1\pm 4\,$km\,s$^{-1}$.}

Chemin discussed the dependence of the observed rotation curve on bar orientation \cite {Chemin2015}.
He used the results of high-resolution hydrodynamical simulations by Renaud et al. ~\cite{Renaud2013}
and demonstrated that the rotation curve of the disk depends on bar orientation relative to an observer. We confirm this result in a more realistic model using combined, direct N-body-gas dynamical simulations.

Deg et al.~\cite{DegGalactICS2019} presented a new version of the GlactICS code, and tested it simulating the dynamics of a Milky Way-like galaxy. The authors used a two-component collisionless-gaseous model, and included bulge and dark matter halo in their simulations. Similar to our results, they find that, after approximately 1 Gyr, a bar formation with semi-axis of about 3 kpc occurs in the disk, and a 
multi-armed spiral pattern, which is more prominent in the gaseous component, is formed outside the bar region.
{Strictly speaking, to model the dynamics of a particular galaxy, one must guess its initial equilibrium, one that fits the final stage of observed equilibrium properties for the galactic disk. However, the disk evolution does not noticeably alter the initial density distribution of the massive stellar disk, and up to 0.5 Gyr does not affect considerably the density distribution of the gaseous disk, thus justifying our approach, as seen in Figure 9. Therefore, the set of unstable spiral modes is well-determined by 
the current state of a slowly evolving disk. The succcessful modeling of spiral structure in nearby spiral galaxies NGC 1566~\cite{Korchagin1566} and NGC 5247 ~\cite{Khoperskov+2012ngc5247}, which was done using the observed equilibrium properties of these systems, further justifies the validity of our approach.}

\subsection{Swing Amplification vs. Global Modes}
In the review paper by Dobbs and Baba~\cite{Dobbs-Baba-2014mod} the authors state "We argue that, with the possible exception of barred galaxies, spiral arms are transient, recurrent, and initiated by swing amplified instabilities in the disc''.  The swing amplification mechanism of spiral formation was suggested by A. Toomre (see, e.g.,~\cite{Toomre1977}) and it remains in many studies the governing mechanism for explaining the origin of spiral structure.
{However, there are} a number of unanswered questions related to this approach. The observed amplitudes of the spirals is one of the oldest of issues with the swing amplification theory. Swing amplification can amplify the initial perturbations by about a hundred times (\cite{Toomre1977}), which is not sufficient for explaining the observed amplitudes of the spiral arms in galaxies.

Similar to other hydrodynamical or plasma systems, galactic disks oscillate, and they have their own global modes that can be unstable. Why are these global modes not considered in the explanation of spiral structure in galaxies?  What happens to these global modes in an explanation involving the swing amplification mechanism?  Why are global modes observed in barred galaxies described as the possible exceptionato the recurrent and swing amplified instabilities? 
Differential rotation and gravity also work in barred galaxies, so the swing amplified noise should be seen in these galaxies too.

It is assumed that the origin of the spiral pattern recently discovered in the galaxy M51a is caused by its satellite galaxy, NGC 5195. However, the closest approach between these two systems was about 900 Myr ago, according to Wahde and Donner~\cite{WihdeDonner2001}. Is the observed two-armed global spiral structure attributed to an interaction that occurred about one Gyr ago?  Why would swing amplification, operating on much shorted time scales, be suppressed in this galaxy?

Sellwood $\&$ Carlberg concluded in their recent paper
\cite{SellwoodCarlberg2020}
"We  argue that apparently shearing transient  spirals in simulations result from the superposition of two or more steadily rotating patterns, each of which is best accounted for as a normal mode of the non-smooth disc.'' 
We completely agree with this statement, and will also cite here Vadim Antonov, the author of classical works in stellar dynamics (Antonov's theorems in Binney and Tremaine's Galactic Dynamics), who said: 
"The galaxies are similar to copper trampets''. 
We can add here that some galaxies, like NGC 1566, "play a pure spiral tone'', while others  "play an accord'' of spiral patterns.

In light of this, a number of questions arise:

\noindent
--- What is the mechanism of spiral saturation at the nonlinear stage of instability? Laughlin et al. 
\cite{LaughlinKorchaginAdams1997} concluded that nonlinear self-interaction of a growing mode is responsible for mode saturation, but further study is needed.

\noindent
--- Why does the presence of gas in the stellar-gaseous gravitating disk extend the lifetime of the spiral pattern? 
Goldreich \& Lynden-Bell~\cite{GoldreichLynden-Bell1965II}
pointed out, back in the sixties, the importance of gas to the formation of spiral structure: S0 galaxies are topographically similar to normal spirals but they have no gas, no dust, and no spiral arms.
The importance of gas in sustaining the spiral structure was also confirmed in numerical simulations
\cite{Khoperskov+2012ngc5247}.
In light of new observational data, the spiral structure in galaxies continues to pose questions to theory.

\section{Summary}

To model the dynamics of a Milky Way-like collisionless-gaseous disk, we use, for the first time in such a study, direct integration in calculating the gravitational potential, contrary to the approximate tree-code or particle-mesh 
codes that were used before. 
It is not clear whether an approximate treatment of the gravity can correctly reproduce the dynamics of the multi-component disks over cosmological time intervals. 
{
An agreement of simulation results in the hydrodynamical, collisionless, and the 
collisionless--hydrodynamical approaches ~\cite{Khoperskov+2012ngc5247} 
provides validity to our model.
}
As for particular results, we demonstrate that:

\noindent
--- An axisymmetric two-component gravitating disk with parameters close to those of the azimuthally averaged Milky Way galaxy---in terms of observed rotation curve, velocity dispersion profile, and masses of the bulge and of the stellar and gaseous disk components---is unstable towards near-exponentially growing spirals having numbers of arms $m =$\,(2--4).

\noindent
--- At the nonlinear stage of instability, the spirals saturate at few tens of percent in the central regions of the disk, to a few percent at the disk's periphery.

\noindent
---  At the nonlinear stage, a prominent bar is formed in the central regions of the disk with a large semi-axis of about 3 kpc.

\noindent
---  Outside the bar region, a complex spiral structure, represented by a superposition of two-, three-, and four-armed spiral patterns, rotating with different angular velocities, is formed. The spiral structure of the Milky Way galaxy is interpreted in some papers as single spiral pattern with a fixed number of arms and spiral pitch angle, and one rotation resonance with fixed position. 
We show that this is not the case for the Milky Way-like disk.

\noindent
---  We demonstrate that the peak in the rotation curve of the disk of the Milky Way galaxy, which is located in its central regions, is a result of the non-circular motions caused by the bar which develops in the disk.

\noindent
---  We confirm that the presence of a massive and centrally concentrated bulge prevents the formation of a bar, in agreement with~\cite{Ostriker-Peebles-1973bar}.

\noindent
---  We also show that the presence of gas of about ten percent the disk's mass extends the lifetime of the spiral structure to a few Gyr as compared to what is found in purely collisionless models. 

{This conclusion though should be validated in further studies with a multi-phase gas model using proper feedback, which is beyond the modelling capabiities of this study.}



\vspace{6pt}



\authorcontributions{Conceptualization, methodology: V.K. and A.K.;
 Writing---original draft preparation, V.K. and A.K.;
 Writing---review \& editing, all authors;
 software and validation, S.K.; numerical simulations, S.K. and A.K.; visualization, S.K. and A.K.; supervision and project administration, V.K.
All authors have read and agreed to the published version of the manuscript.}


\funding{The development of software for modeling galaxies and numerical experiments were funded by the Ministry of Science and Higher Education of the Russian Federation (the government task no. 0633-2020-0003, S.K. and A.K.).}
\institutionalreview{\hl{please add}}

\informedconsent{\hl{please add}}


\dataavailability{\hl{please add}} 
\acknowledgments{Authors thank T. Girard and the referees for valuable comments. Numerical simulations were carried by using the equipment of the shared research facilities of ``Supercomputer 
Center of Volgograd State University''. V.K.  acknowledges  financial support by Southern Federal University,  2020 (Ministry of 
Science and Higher Education of Russian Federation).}

\conflictsofinterest{The authors declare no conflict of interest.}

\abbreviations{The following abbreviations are used in this manuscript:\\

\noindent
\begin{tabular}{@{}ll}
COBE / DIRBE  & Cosmic Background Explorer / Diffuse Infrared Background Experiment \\
SPH & Smoothed-particle hydrodynamics \\
OpenMP-CUDA & Open Multi-Processing - Compute Unified Device Architecture \\
VLBI & Very Long Baseline Interferometry
\end{tabular}}%



\end{paracol}
\reftitle{References}

\end{document}